\documentclass[fleqn,10pt]{wlscirep}

\usepackage{pslatex,graphicx,dcolumn,bm,amssymb,amsmath,color,mathtools}

\usepackage[utf8]{inputenc}
\usepackage[T1]{fontenc}
\usepackage{soul}
\usepackage{comment}
\usepackage{multirow}
\usepackage{ulem}
\usepackage{hyperref}
\usepackage{cancel}

\usepackage{xcolor}

\newcommand{\be}{\begin{equation}}
\newcommand{\ee}{\end{equation}}

\DeclareMathOperator*{\argmin}{arg\,min}

\begin{document}

\title{Tunable Hyperuniformity and Hidden Information in Random  Cellular Structures}

\author[1,2]{Yiwen Tang}
\author[1]{Mostafa Alkady}
\author[3]{Xinzhi Li}
\author[1,*]{Dapeng Bi}

\affil[1]{Department of Physics and the Center for Theoretical Biological Physics, Northeastern University, Boston, MA 02115, USA}
\affil[2]{Mechanobiology Institute, National University of Singapore, Singapore}
\affil[3]{School of Physics, Georgia Institute of Technology, Atlanta, GA 30332, USA}

\affil[*]{Corresponding author: d.bi@northeastern.edu}

\begin{abstract}
Hyperuniform systems possess the isotropic microstructure of liquids while suppressing large-scale density fluctuations with crystal-like precision, endowing them with exotic optical and transport properties. However, generating these states typically relies on top-down algorithmic optimizations that lack physical realism, leaving the fundamental mechanical and thermodynamic limits of such disordered order largely unexplored. Here, we utilize a mechanical vertex model to
explore physical conditions that drive self-organized hyperuniformity in this model. By tuning cortical elasticity and interfacial tension, we engineer a continuous spectrum of effectively hyperuniform states.  Crucially, hyperuniformity and rigidity are independently tunable in this system: hyperuniform states occur on both sides of the solid--fluid transition, and the degree of hyperuniformity is set by cell mechanics rather than by the onset of rigidity. Building on these states, we introduce the Hyperuniform Poisson Ensemble (HyPE), constructed by overlaying independent hyperuniform subsets. HyPE states preserve long-range hidden order while asymptotically approaching Poissonian randomness at short length scales, establishing universal scaling relations for density fluctuations. As the most disordered hyperuniform systems reported, HyPEs exhibit vanishing configurational entropy, establishing a fundamental thermodynamic lower bound for the information content of hyperuniform matter. By bridging cellular mechanics with the statistical physics of hidden information, this framework provides a universal blueprint for designing multifunctional metamaterials, with direct applications in photonic bandgap engineering, phononic control, and stress-responsive composites.

\end{abstract}

\maketitle

Density fluctuations offer a powerful lens for classifying the structural order of many-particle systems~\cite{vezzetti1975new,ziff1977bulk,jiao2011,wax2002}. While crystalline solids exhibit long-range periodic order and fluids display large-scale density variations, hyperuniform materials occupy a unique middle ground. They are isotropic and disordered like liquids, yet they suppress large-scale density fluctuations with crystal-like precision~\cite{torquato2003,torquato2018,klatt2019,fernandez2024hyperuniform, wilken2023spatial}. Formally, this "hidden order" is captured by a vanishing structure factor in the infinite-wavelength limit, $\lim_{q\rightarrow0}S(q)=0$~\cite{torquato2003}. This exotic duality endows hyperuniform systems with properties traditionally reserved for crystals, including widely tunable optical~\cite{leseur2016high,florescu2009designer,li2018a,froufe2016role}, acoustic~\cite{gkantzounis2017hyperuniform,cheron2022experimental}, mechanical~\cite{xu2017microstructure,chen2018designing,torquato2000modeling,zhang2016perfect,li2023_sm}, and structural~\cite{salvalaglio2024persistent,Puig_PRB_2024} characteristics.

Biological tissues provide a natural, mechanically driven platform to address this gap. Often amorphous yet spatially regulated, tissues must balance macroscopic mechanical fluidity with the need for uniform spatial coverage~\cite{atiaGeometricConstraintsEpithelial2018,bi2016,garcia2015,mongera2018,yang2021,tang2025packing,kayal2026}. Recent research suggests that hyperuniformity may serve as a fundamental metric for tissue organization~\cite{li2018a,zheng2020}, with experimental evidence of "true" hyperuniformity observed in avian photoreceptor patterns~\cite{jiao2014}. However, a comprehensive framework connecting single-cell mechanics—such as cortical tension and adhesion—directly to the emergence of macroscopic hyperuniform order has remained elusive.

Here, we introduce a theoretical framework where hyperuniformity arises naturally from the interplay of local mechanics and geometric constraints. Using a Self-Propelled Voronoi (SPV) model, we identify the specific mechanical conditions—governed by the competition between cortical elasticity and interfacial tension—that drive self-organized hyperuniformity.   Disordered hyperuniform states are well documented in both rigid and flowing matter, from jammed packings and amorphous solids~\cite{torquato2003,klatt2019} to one-component plasmas and polymer melts~\cite{lebowitz1983,lomba2017,chremos2018}. What our model adds is that both arise within a single system from the same microscopic mechanics: hyperuniformity and rigidity can be tuned independently, so that the tissue can be made effectively hyperuniform on either side of the solid--fluid transition, with the degree of hyperuniformity controlled by cortical elasticity rather than by the onset of rigidity.

Finally, we generalize these findings to propose a new class of disordered matter: the \textbf{Hy}peruniform \textbf{P}oisson \textbf{E}nsemble (HyPE). Constructed by overlaying independent hyperuniform subsets, HyPE states preserve long-range hidden order while asymptotically approaching Poissonian randomness at short scales. By decoupling long-range order from local geometric configurations, HyPE states exhibit vanishing configurational entropy, establishing a fundamental thermodynamic lower bound for the information content of hyperuniform systems. This mechanical and information-theoretic framework not only advances our understanding of biological self-organization but also provides a universal blueprint for designing multifunctional metamaterials with tunable fluctuation amplitudes.

\section*{Model}

Epithelial cells, when densely packed into a monolayer, form a space-filling polygonal tiling. The geometric organization of this tiling is shaped by the mechanics of cell-cell interactions and the cells' interactions with their environment,
effectively tuning the degrees of both short and long-range orders within the tissue. This makes the cellular structures excellent candidates for studying hyperuniformity.

In computational studies, vertex-based models have been extensively employed to analyze 2D tissue monolayers and have shown remarkable agreements with experimental data~\cite{farhadifar2007, park2015, atiaGeometricConstraintsEpithelial2018, mitchelPrimaryAirwayEpithelial2020}. One such model, the Self-Propelled Voronoi (SPV) model~\cite{bi2015, bi2016}, utilizes Voronoi tessellation~\cite{dirichlet1850, honda1978} based on the cell centers $\{\vec{r}_i\}$ to define cell shapes and the connective topology of the cellular network. This approach provides a robust framework for exploring the emergent properties of cellular patterns and their potential for hyperuniform order.  When all single-cell properties are equal,
the non-dimensionalized form of the tissue's mechanical energy\cite{bi2015} is given by
\begin{equation}
    \epsilon = \sum_{i=1}^N \left[(a_i-1)^2+K(p_i-p_0)^2\right],
    \label{eq:epsilon}
\end{equation}
where $a_i$ and $p_i$ denote the associated rescaled cell area and perimeter, respectively. In this formulation, the preferred area is set to be $1$.
In Eq.~\eqref{eq:epsilon}, the term quadratic in cell areas results from cell incompressibility and the monolayer's resistance to height fluctuations\cite{farhadifar2007,staple2010a, hufnagel2007, zehnder2015}.
The term for cell perimeters arises from the contractility of the cell cortex.
The cortical elasticity $K$ reflects relative contributions to the energy functional from deviations in cell perimeter and area from their preferred values \cite{farhadifar2007}.
The target shape index $p_0$, defined as the ratio of the target perimeter to the square root of the target area, modulates the balance between cell–cell adhesion and cortical contractility, thereby governing the solid–fluid transition in the tissue.
Throughout, ``solid'' and ``fluid'' refer to the mechanical response measured by the shear modulus, $G>0$ and $G=0$ respectively (see Methods). This transition is often termed jamming/unjamming in the epithelial literature~\cite{park2015,atiaGeometricConstraintsEpithelial2018,bi2016}; we use the mechanical terminology here because the tissue is confluent at all times and the rigidity transition is density-independent, controlled by $p_0$ rather than by crowding~\cite{bi2015,arora2024}.
 Two key parameters, $K$ and $p_0$, are the focus of our study conducted through computer simulations using the open-source cellGPU~\cite{sussman2017} (see Methods).
The mechanically stable states of the SPV model are achieved via a fast inertial relaxation engine~(FIRE) algorithm~\cite{FIRE_PRL_2006,guenoleAssessmentOptimizationFast2020} under periodic boundary conditions.
All configurations studied in this work are athermal: the simulations contain neither thermal noise nor self-propulsion, and every configuration is a force-balanced local minimum of Eq.~\eqref{eq:epsilon}, i.e. an inherent structure in the sense standard to the study of amorphous solids~\cite{stillinger1982hidden}. All mechanical characterization reported below is obtained from the linear response of these same minima (see Methods); we therefore make no claims regarding dynamics.

\begin{figure*}[ht]
    \centering
    \includegraphics[width=0.9\textwidth]{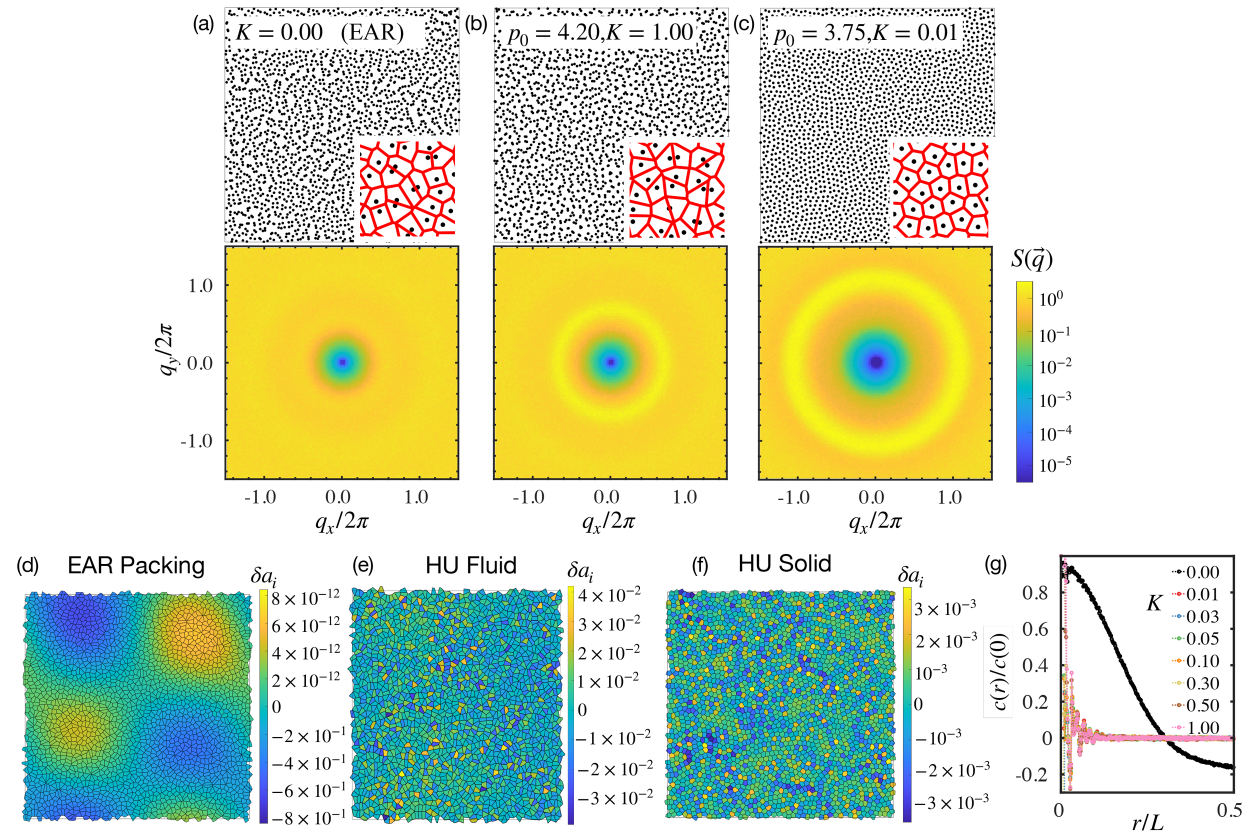}
    \caption{
    \textbf{Hyperuniformity in model biological multicellular structures.} {\bf Top row}: Snapshots for EAR-packing with  $K=0.00$ {\bf(a)}, hyperuniform fluid state with $p_0=4.20, K=1.00$ {\bf(b)}, and hyperuniform solid state with $p_0=3.75, K=0.01$ {\bf(c)}  and their corresponding 2D structure factor $S(\vec{q})$.  The insets depict a magnified view of cell centers and corresponding Voronoi tessellations.
    Snapshots for  EAR packing with $K=0.00$ {\bf(d)} and hyperuniform fluid state with $p_0=4.2, K=1$ {\bf(e)} and hyperuniform solid state with $p_0=3.75, K=0.01$ {\bf(f)}, where cells are colored by their area deviations from the mean, $\delta a_i=a_i-\langle a\rangle$. {\bf(g)} The real-space area correlation function as a function of $r$ for EAR Packing ($K=0.00$) vs that for solid states ($p_0=3.75, K=0.01, 0.03, 0.05, 0.10, 0.30, 0.50, 1.00$).
    }
    \label{fig:hu_states_overview}
\end{figure*}

\section*{Results}

\paragraph*{\textbf{Equal-Area Random (EAR) Packings}}
The original definition of hyperuniformity, which refers to suppressed density~\cite{torquato2003,gabrielli2008} or  {\bf equivalently, area fluctuations}, inspires a minimal way to achieve hyperuniformity in the model tissue collective: {\it enforcing uniform cell areas in the polygonal tiling}.
The tissue energy (Eq.~\ref{eq:epsilon}) naturally includes such an area-constraining term, and setting $K=0$ enforces this constraint exactly.
When the system reaches the ground state (i.e. $\epsilon = 0$), it is by definition hyperuniform.
Equal area is the only condition imposed: at $K=0$ the perimeter term drops out of Eq.~\eqref{eq:epsilon}, so every configuration of cell centers whose Voronoi cells have equal area is an EAR packing, and no constraint is placed on cell shape or topology. Packings satisfying any additional condition, including the finite-$K$ states considered below, are not EAR packings even when their areas are nearly uniform.
Furthermore, these ground states are highly degenerate and continuous ~\cite{bi2016}, allowing the system to explore a vast configuration space without energetic penalties. We refer to these   Voronoi-based Equal Area Random packings as EAR packings. An example is shown in Fig.~\ref{fig:hu_states_overview}(a) along with the ensemble-averaged structure factor $S(\vec{q})$ , where $\vec{q}$ is the wavevector (see Methods), which decreases to well below $10^{-4}$ near $\vec{q} = 0$.
EAR packings are strictly hyperuniform ~\cite{godfrey2018absence,puig2022anisotropic}, evidenced by the continuous decrease of the angularly averaged $S(q)$, where $q=|\vec{q}|$, as $q$ diminishes without reaching saturation, $S(q)\sim q^4$~(Fig.~S1(a)).
EAR packing is a simple yet powerful way to generate hyperuniformity, which has not been fully explored as a natural route to generating diverse hyperuniform configurations.
In the following section, we will systematically explore the diverse classes of hyperuniform states achievable by manipulating $K$ and $p_0$. Throughout our work, EAR packing serves as a baseline for comparison with other hyperuniform structures.

\begin{figure}
    \centering
    \includegraphics[width=0.75\columnwidth]{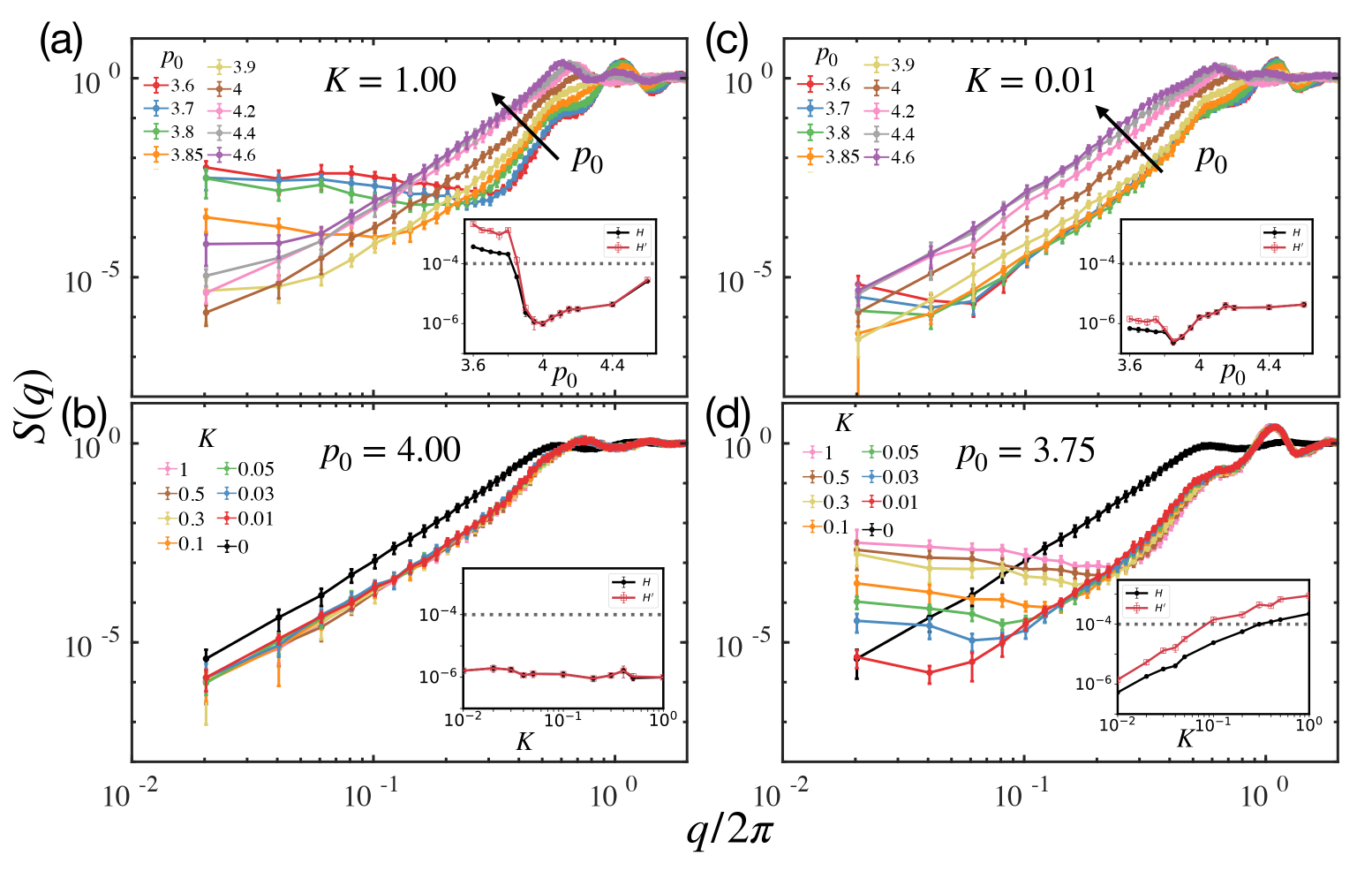}
     \caption{
     \textbf{ Characterization of long-range order using the structure factor $S(q)$ and the hyperuniformity index $H$ under  distinct mechanical regimes}:
     {\bf (a)} When the cell perimeter is strongly constrained  ($K=1.00$, varying $p_0$);
     {\bf (b)} In the HU fluid phase ($p_0=4.00$, varying $K$);
     {\bf (c)} When the cell perimeter is weakly constrained ($K=0.01$, varying $p_0$);
     {\bf (d)} In the solid phase  ($p_0=3.75$, varying $K$).
     The insets show $H \equiv \min_{q} S(q)/S(q_{peak})$ and $H' \equiv S(q_{\rm min})/S(q_{peak})$ (see Methods) as a function of $p_0$ in (a,c)  and as a function of $K$ in (b,d). The black dashed lines denote the empirical threshold of $H = 10^{-4}$, commonly used to define hyperuniformity in finite systems. All data correspond to simulations with $N=2430$ cells. Error bars represent the standard deviation over different initial seeds.}
     \label{fig:sq_regimes}
\end{figure}

\paragraph*{\textbf{Hyperuniform Fluid State}}
In vertex-based models, extensive research has demonstrated that the target shape index $p_0$ governs a mechanical phase transition between solid and fluid states of biological tissues—solid when $p_0 < p_0^*$ and fluid for $p_0 > p_0^*$~\cite{bi2015,yan2019,das2021}. Further studies indicate potential hyperuniformity in the fluid phase~\cite{li2018a,zheng2020}. However, these investigations were constrained by limited system sizes~\cite{li2018a} and provided inconclusive results regarding whether the fluid phase is consistently hyperuniform or only exhibits hyperuniformity at specific values of $p_0$~\cite{zheng2020}. Therefore, we systematically examine how hyperuniformity might manifest across different phases of the SPV model.

While maintaining a constant cortical elasticity of $K=1$, we explored target shape index $p_0 \in [3.60, 4.60]$, spanning across the critical rigidity point at $p_0^*\approx3.813$ \cite{yan2019},  thereby encompassing both solid and fluid states. The structure factor $S(q)$ for these $p_0$ values shows suppressed density fluctuations as $q\rightarrow 0$, as confirmed in Fig.~\ref{fig:sq_regimes}(a).
We adopt the empirical criterion that the hyperuniformity index $H$~\cite{torquato2018} $< 10^{-4}$ serves as the threshold for hyperuniformity, following Ref.~\cite{atkinson2016} (see Methods).
We observed that a subset of the solid phase ($p_0<p_0^*$) exhibits non-hyperuniformity, with $H$ values typically above $10^{-4}$. At $p_0^*$, $H$ sharply drops to around $10^{-6}$, indicating consistent hyperuniformity in the fluid phase.
While strictly vanishing density fluctuations are only possible at $K=0$, fluid states with finite $K$ exhibit a low-$q$ plateau in $S(q)$, indicating effective hyperuniformity at biologically relevant length scales~(Fig.S1(b)). This plateau prevents strict hyperuniformity, further distinguishing the fluid from EAR packings through systematic variations in $K$ (Fig.~\ref{fig:sq_regimes}(b)).
Accordingly, the EAR packings at $K=0$ are the only strictly hyperuniform states in this work. All states with $K>0$, both fluid and solid, are \emph{effectively} hyperuniform in the sense of the criterion $H<10^{-4}$ defined in Methods. For brevity we use ``hyperuniform'' as shorthand for   ``effectively hyperuniform'' in what follows, reserving ``strictly hyperuniform'' for the $K=0$ case.

\paragraph*{\textbf{Hyperuniform Solid State}}
EAR-packing is only constrained by cell area and is inherently hyperuniform. However, the introduction of constraints on cell perimeters counteracts the hyperuniformity in the solid phase, as shown in Fig.~\ref{fig:sq_regimes}(a).
This prompts the question: Can we tilt the balance in favor of the cell area constraint,  thereby fostering hyperuniformity in the solid phase? Fig.~\ref{fig:sq_regimes}(c) suggests that a reduction in $K$ could achieve it, giving  $H\ll 10^{-4}$. The perimeter constraint merely obscures the hyperuniformity that the solid phase is capable of sustaining.
A representative example of a hyperuniform solid state is shown in Fig.~\ref{fig:hu_states_overview}(c). It possesses $S(\vec{q}\rightarrow 0)$ values well below $10^{-4}$ but displays unique configurations and peaks in structure factor, which distinguishes it from EAR-packing and hyperuniform fluid states.
Fig.~\ref{fig:sq_regimes}(d) illustrates how $K$ affects the structure factor of a solid state at fixed $p_0=3.75$. Decreasing $K$ shifts the low-$q$ plateau in $S(q)$ downwards.  We further demonstrate that the plateau remains unchanged across different system sizes and initial conditions (Fig.~S1(d) and Fig.~S2).
Consequently, the hyperuniformity index $H$ descends below a threshold $10^{-4}$ corresponding to hyperuniformity, despite being in the solid state.
This suggests that the hyperuniformity is significantly influenced by the cortical elasticity $K$ and it does not always coincide with solid or fluid behavior which is fully characterized by the shear and bulk moduli (Fig.~S3). Therefore, hyperuniformity and rigidity can be independently tuned.

To further differentiate hyperuniform solid and fluid states from EAR packings, we analyze the spatial distribution of cell areas. Fig.~\ref{fig:hu_states_overview}(d-f) illustrates the distinct area distributions across these states.
A key feature emerges in the normalized area spatial correlation function $c(r)/c(0)$, where $c(r) = \langle \delta a_i(r) \delta a_j(0) \rangle$ and $\delta a_i = a_i - \langle a \rangle$.
In EAR packings, the area correlation length spans the size of the system, whereas in both hyperuniform solid and fluid states, the correlations decay rapidly at short distances.  This highlights that, although hyperuniformity and rigidity can be independently tuned, their interplay is further shaped by local structural organizations, reinforcing that hyperuniformity alone is insufficient to fully characterize the distinctions between these phases.

\paragraph*{\textbf{Short-Range Order}}

\begin{figure*}
    \centering
    \includegraphics[width=0.75\textwidth]{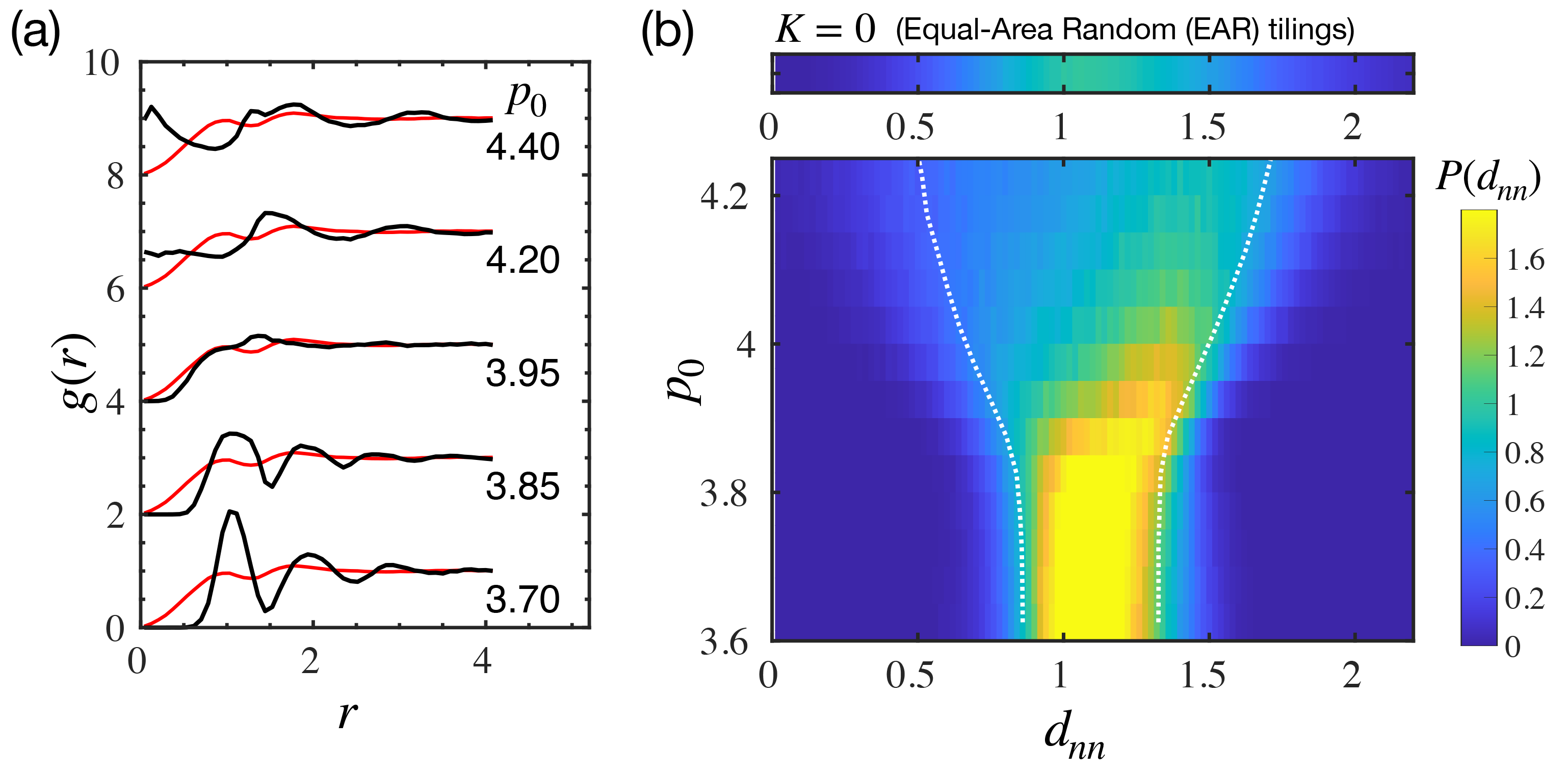}
    \caption{
    \textbf{Increased $p_0$ weakens the short-range order.}
    {\bf (a)} The pair-correlation function $g(r)$ at different values of $p_0$ for a system of size $N = 2430$. Distances are measured in units of the average cell-cell separation, which serves as the unit of length throughout. The red line represents $g(r)$ in EAR-packing ($K=0$) for comparison.  The values of $g(r)$ have been shifted for visual clarity.
    {\bf (b)} Probability density function of nearest-neighbor distances $d_{\mathrm{nn}}$ between cells for various $p_0$ values and EAR packings. The width of each distribution is highlighted by white dashed lines, which mark $\langle d_{\mathrm{nn}}^{\min}\rangle$ and $\langle d_{\mathrm{nn}}^{\max}\rangle$, the averages over all cells of the smallest and largest $d_{\mathrm{nn}}$ from a cell to its adjacent neighbors.
}
    \label{fig:short_range_order}
\end{figure*}

In addition to long-range order, the model tissue can possess a large variation of short-range order. We compute the pair correlation function $g(r)$ as a function of the radial distance $r$ (see Methods) for varying values of $p_0$ at $K=1$~(Fig.~\ref{fig:short_range_order}(a)).
For the solid phase ($p_0<3.81$),   there is strong mutual exclusion between nearest neighbors due to effective short-range repulsion~\cite{bi2016}, thereby suppressing the pair correlation function  $g(r)$ at short distances up to the first peak $r \approx 1.1$.\footnote{The packing of regular hexagons with a unit area has a nearest-neighbor distance of approximately $d_{\mathrm{nn}} = \sqrt{2/\sqrt{3}} \approx 1.075$.}
At longer distances, the peaks of $g(r)$ decay rapidly, similar to a fluid~\cite{dijkstra2000simulation}.
In the fluid phase ($p_0>3.81$), the peaks of $g(r)$ broaden and become increasingly indistinguishable at higher $p_0$'s. There is also a lack of neighbor-neighbor exclusion as the cells no longer interact via purely repulsive interactions~\cite{bi2016}. Deeper into the fluid regime ($p_0>4.20$), cells can become arbitrarily close. Consequently, the first peak in $g(r)$ approaches $r=0$, leading to a loss of short-range order in the system—making it resemble a gas rather than a fluid.

Since  $g(r)$ is a distance-based metric, it loses sensitivity to short-range order when its peaks become indistinct. To complement this, we also analyze a topology-based metric, $P(d_{\mathrm{nn}})$, where $d_{\mathrm{nn}}$ is the distance between the centers of adjacent cells and the subscript denotes nearest neighbor (Fig.~\ref{fig:short_range_order}(b)) (as determined by the Voronoi tesselation). In the solid phase, $P(d_{\mathrm{nn}})$ has a narrow peak centered at around $d_{\mathrm{nn}} \approx 1.1$, corresponding to the first peak in $g(r)$. However, in the fluid phase, as $p_0$ is increased, $P(d_{\mathrm{nn}})$ broadens significantly as its mean increases.
This broadening arises because cells elongate to accommodate higher target perimeters, increasing the variability in distances between neighboring cells. When two elongated neighboring cells align, their centers move closer together, whereas non-aligned pairs exhibit greater separations.
These two types of nearest-neighbor pairs are illustrated in Fig.~\ref{fig:short_range_order}(b) by the dotted white lines. For each cell we take the smallest and the largest of the distances $d_{\mathrm{nn}}$ to its adjacent neighbors, $d_{\mathrm{nn}}^{\min}$ and $d_{\mathrm{nn}}^{\max}$; the two lines mark the averages $\langle d_{\mathrm{nn}}^{\min}\rangle$ and $\langle d_{\mathrm{nn}}^{\max}\rangle$ over all cells and realizations.
Our short-range analysis is performed at $K = 1$, but the findings hold across the range of $K$ studied here. Although $K$ strongly affects the long-range order, as seen in $S(q)$, short-range order is essentially unchanged for $K \gtrsim 10^{-5}$ and varies smoothly and continuously below this value, approaching the EAR-packing result as $K$ is reduced further (Fig.~S4). The EAR-packing case ($K=0$) nonetheless remains distinguishable from this limit: at $K=0$ every equal-area tiling is an exact ground state, so the ground states are degenerate, whereas any $K>0$ lifts this degeneracy and selects those equal-area configurations that additionally minimize the perimeter term. The distinct $g(r)$ and $P(d_{\mathrm{nn}})$ of EAR packings (Fig.~\ref{fig:short_range_order}) therefore reflect this degeneracy rather than a transition at $K=0$.

We also examined bond-orientational order~\cite{tang2024hexatic}. The two-dimensional $S(\vec{q})$ maps show no directional bias, indicating that these configurations are structurally isotropic, and the distributions of the global order parameters $\Psi_2$, $\Psi_4$ and $\Psi_6$, together with the local $\psi_6$, confirm that no significant orientational order is present in any of the four ensembles (Fig.~S11).

\paragraph*{\textbf{ A phase diagram linking hyperuniformity and mechanics } }

\begin{figure}[h]
    \centering
    \includegraphics[width=0.5\columnwidth]{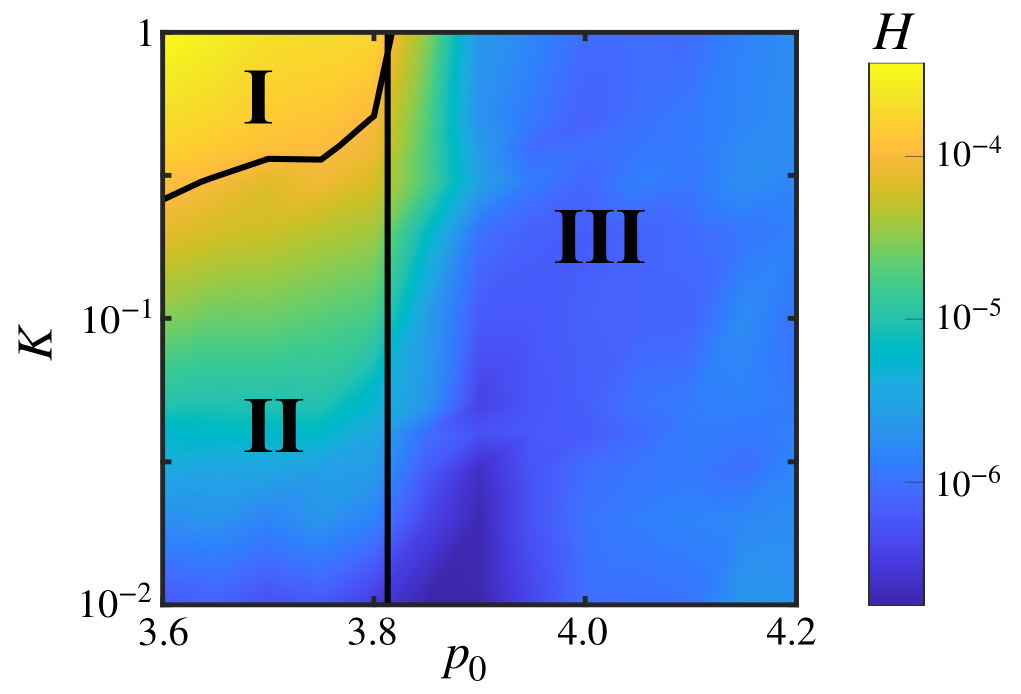}
    \caption{\textbf{Phase diagram of hyperuniformity index $H$ as a function of target shape index $p_0$ and {cortical elasticity} $K$.} It encompasses three distinctive phases: (I) non-hyperuniform solid state, (II) hyperuniform solid state, and (III) hyperuniform fluid state. Boundary curve separating regions I and II corresponds to $H=10^{-4}$, while line separating II and III corresponds to $p_0=3.81$. Representative configurations and structure factors for each region are shown in Fig.~S10.}
    \label{fig:phase_diagram}
\end{figure}

The properties of hyperuniformity and rigidity define three different phases quantified by the hyperuniformity index and elastic modulus. The results are summarized in a phase diagram in Fig.~\ref{fig:phase_diagram}.
The boundary between regions I and II is the $H=10^{-4}$ contour, while that between II and III is the rigidity transition at $p_0=3.81$; the former is a criterion applied to a continuously varying index, the latter a mechanical transition at which $G$ vanishes.
In the solid regime $p_0<3.81$, tissues are all solid but exhibit a transition from hyperuniformity to non-hyperuniformity with increasing $K$. In the fluid regime $p_0>3.81$, tissues are all fluid and hyperuniform.
Representative configurations for each of the three regions are shown in Fig.~S10: (I) $p_0=3.7$, $K=0.8$; (II) $p_0=3.7$, $K=0.03$; and (III) $p_0=4.0$, $K=0.1$. Regions I and II share the same target shape index and differ only in cortical elasticity, so their cell shapes are similar; the distinction between them is not apparent locally but emerges at long wavelengths(as also reflected in Fig.~\ref{fig:sq_regimes}(d)). Region III, at larger $p_0$, contains visibly more elongated and irregular cells.

Having examined long- and short-range orders on the tissue scale, distinguishing EAR packings from
hyperuniform fluid states, we next sought the potential single-cell level mechanisms that underlie
these different hyperuniform behaviors. Here, we analyze the distribution of cell morphologies at
various $K$ and $p_0$ values, as shown in Fig.~S5.

Three statistics of the cell population control the energetics: the standard deviation of cell
areas, $\sigma_a$; the deviation of the mean cell perimeter from its target,
$\delta p \equiv \langle p_i \rangle - p_0$; and the standard deviation of cell perimeters,
$\sigma_p$. Because the mean cell area is unity by construction, the mean energy per cell of
Eq.~\eqref{eq:epsilon} separates exactly into these three contributions,
\begin{equation}
    \frac{\langle \epsilon \rangle}{N} \;=\; \sigma_a^2 \;+\; K\,\delta p^{\,2} \;+\; K\,\sigma_p^2 .
    \label{eq:energy_decomposition}
\end{equation}
The first term penalizes heterogeneity in cell area, the second a systematic mismatch between
the typical cell perimeter and its target, and the third heterogeneity in cell perimeter.
Minimizing the energy therefore drives the tissue down a gradient with three competing
contributions,
\begin{equation}
    -\frac{\partial}{\partial r}\frac{\langle \epsilon \rangle}{N}
    \;=\; -\left(
    2\sigma_a \frac{d\sigma_a}{dr}
    \;+\; 2K\,\delta p\, \frac{d\,\delta p}{dr}
    \;+\; 2K\sigma_p \frac{d\sigma_p}{dr}
    \right),
    \label{eq:energy_gradient}
\end{equation}
where $r$ denotes the cell-center coordinates collectively. Which of the three the minimization
actually reduces is set by their relative magnitudes, and this is what distinguishes the solid and
fluid regimes.

In the solid phase ($p_0 < 3.81$), the average cell perimeters $\langle p_i \rangle$ hover around
$3.81$~\cite{bi2016,malakar2023}, resulting in significant deviations
$\delta p = \langle p_i - p_0 \rangle > \sigma_p$. The solid transition point is approximately where
fluctuations begin to dominate over deviations.
At $K=1$, the large $\delta p$ therefore makes the second term of
Eq.~\eqref{eq:energy_decomposition} the dominant one.
Larger values of $K$ generally aim to reduce $\delta p$ but weaken area constraints, resulting in
higher values of $\sigma_a$ and thereby preventing hyperuniformity.

In contrast, within the fluid state, $\langle p_i \rangle$ demonstrates a linear growth pattern with
$p_0$, keeping $\delta p < \sigma_p$.
The second term of Eq.~\eqref{eq:energy_decomposition} is then small however large $K$ is, so
the minimization acts principally on $\sigma_a$.
This small deviation $\delta p$ enables effective suppression of fluctuations, allowing the
emergence of hyperuniform fluid states even at large $K$.
This is consistent with $\sigma_a$, rather than $K$ or $p_0$ individually, being the quantity
that tracks the hyperuniformity index (Fig.~S6).
The logarithmic scale of $\delta p$ reveals a valley located within the range of $[4.10,4.20]$
(Fig.~S5(a)).

Decreased $K$ corresponds to weaker perimeter control,  having higher deviation $\delta p$ and larger fluctuation $\sigma_p$, while exerting stronger control over the cell area, resulting in smaller fluctuation $\sigma_a$.
The fundamental characteristic underlying hyperuniformity lies in the standard deviation of cell area $\sigma_a$, which correlates well with $H$, as shown in Fig.~S6.

Both $\sigma_p$ and $\sigma_a$ reach their minima at $p_0 \approx 4$, which is close to the average cell perimeter $\langle p_i \rangle = 4.08$ observed in EAR packings. The slight discrepancy between 4 and 4.08 could potentially be attributed to the asymmetric distribution of $p_i$ in EAR packing compared to the Gaussian constraint in finite $K$.

\subsection*{The Hyperuniform Poisson Ensemble (HyPE)}

\begin{figure*}
    \centering
    \includegraphics[width=1\textwidth]{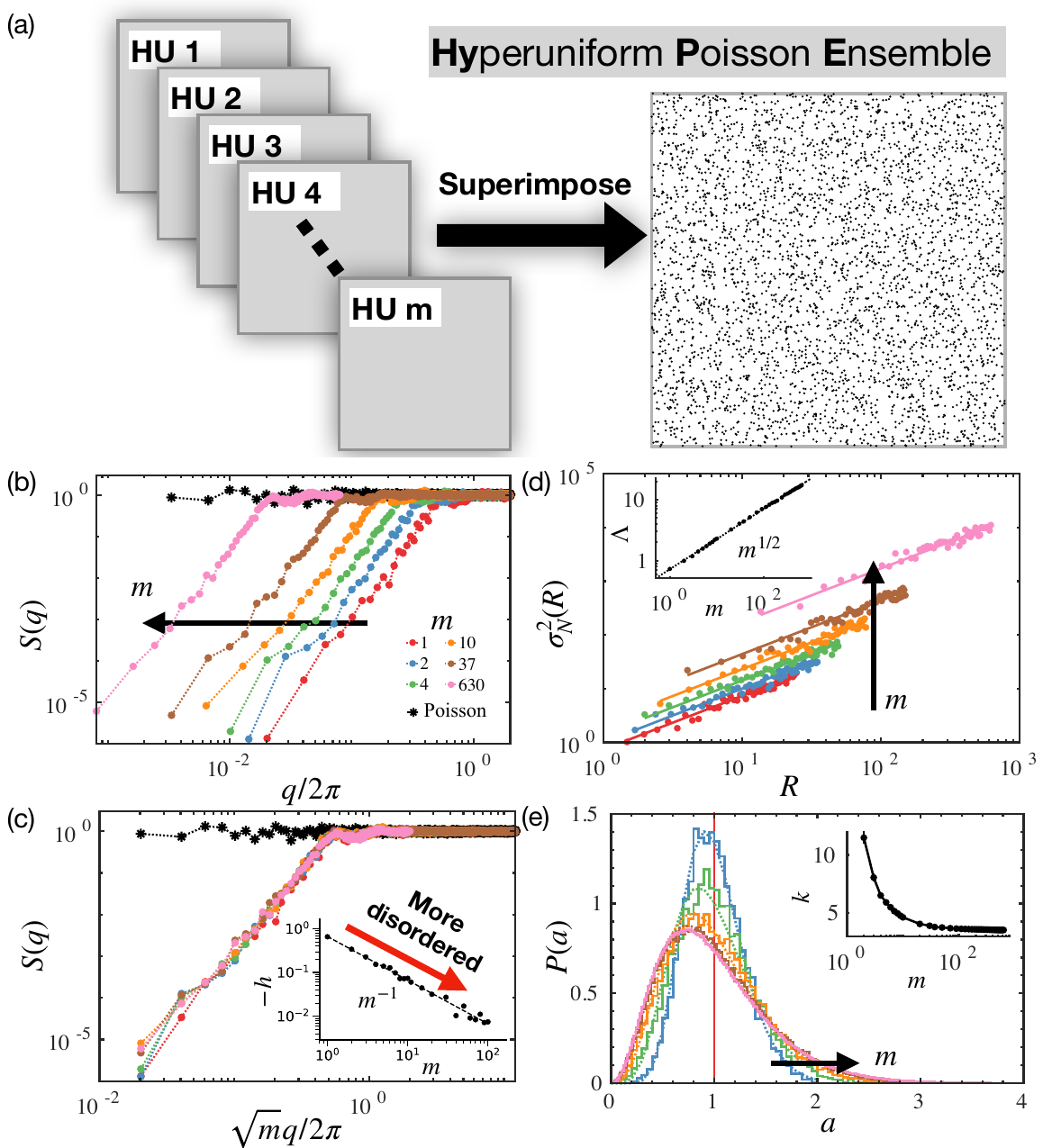}
    \caption{\textbf{Properties of  {\bf Hy}peruniform {\bf P}oisson {\bf E}nsemble (HyPE) based on EAR packings.}
    {\bf (a)} Schematic for generating HyPE states.
    {\bf (b)}  Structure factor $S(q)$ vs. $q$ for various values of $m$ can be rescaled to form a scaling collapse when plotted against $\sqrt{m} q / 2\pi$  {\bf (c)}, indicating self-similarity of HyPE states in $q$-space. The entropy follows a scaling relation $-h\propto m^{-1}$ (inset).
    {\bf (d)}  Density fluctuation of HyPE states obey $\sigma_N^2(R) = \Lambda(m) R$, where the associated surface constant obeys the scaling relation  $\Lambda \propto \sqrt{m}$ (inset).
    {\bf (e)}  Cell area distributions $P(a)$ in HyPE states are well-described by $\Gamma$ distributions $P(a) = a^{k-1} e^{-a} / \Gamma(k)$ (dashed lines) with the shape parameter $k(m)$ shown in the inset.
    }
    \label{fig:hype_properties}
\end{figure*}

Controlling the area variation $\sigma_a^2$ is critical for achieving hyperuniform states derived directly from our Voronoi models, as it correlates with the hyperuniformity index $H$ (Fig.~S6). However, it remains unclear whether low area variance is strictly necessary for hyperuniformity, or if this constraint can be relaxed. By definition, hyperuniformity reflects suppressed density fluctuations at long wavelengths, and is therefore relatively insensitive to short-range structural details. The diverse short-range ordering observed in Fig.~\ref{fig:short_range_order} supports the possibility that hyperuniformity can coexist with variable local area distributions.

To directly test this, we introduce the \textbf{Hy}peruniform \textbf{P}oisson \textbf{E}nsemble (HyPE) states. As illustrated in Fig.~\ref{fig:hype_properties}(a),
a HyPE state is constructed by taking the union of $m$ statistically independent hyperuniform point sets, each of $N$ points at unit density in a periodic box of side $L=\sqrt{N}$, and then dilating all coordinates by $\sqrt{m}$ so that the merged pattern of $mN$ points again has unit density.
Superpositions of this kind occur physically: avian photoreceptor mosaics are multihyperuniform, each cone type forming a hyperuniform pattern whose union is also hyperuniform~\cite{jiao2014}, and multihyperuniformity has recently been reported in high-entropy MXenes~\cite{liu2025}. HyPE is the limiting case in which the subsets are statistically independent, and therefore bounds how disordered a hyperuniform pattern can be at short range.
Fig.~\ref{fig:hype_properties}(b) shows that this superposition preserves the long-range scaling of the structure factor, $\lim_{q\to 0} S(q)   \propto q^4$. This means the overlapping of hyperuniform patterns does not disrupt hyperuniformity at large length scales. However,  it simultaneously suppresses short-range order rapidly when the number $m$ of superimposed point sets is increased, with $S(q)$ approaching that of a Poisson Point Pattern (PPP)~\cite{brakke1987statistics} for $m \to \infty$.  In Fig.~S7, we also show that $g(r)$ rapidly suppresses its peak and converges to the PPP with increasing $m$.

\paragraph*{\textbf{ HyPE states exhibit  self-similarity } }
Interestingly, the value of $q$ at which $S(q)$ deviates from $q^4$ scaling (and flattens with $q$) depends systematically on $m$. In Fig.~\ref{fig:hype_properties}(c), we replot $S(q)$ as a function of $\sqrt{m}q$ and find that all curves collapse onto a single master curve, suggesting that HyPE states at different $m$ are in fact  self-similar.
Both the preservation of hyperuniformity and this collapse follow directly from the construction. Writing $A_s(\vec q)=\sum_{j\in s}e^{-i\vec q\cdot\vec r_j}$ for the contribution of subset $s$, the structure factor of the union is $\big\langle \big| \sum_s A_s \big|^2 \big\rangle/(mN)$. Because the subsets are independent and each is statistically translation invariant, $\langle A_s A_t^{*}\rangle = 0$ for $s\neq t$ and $\vec q \neq 0$; the cross terms vanish and the union has the same structure factor as a single subset. The subsequent dilation by $\sqrt{m}$ maps $\vec q \to \sqrt{m}\,\vec q$, so that
\be
    S_m(q) = S_1(q \sqrt{m}).
\label{m-scaling}
\ee
Equation~\eqref{m-scaling} is exact. It accounts for the collapse in Fig.~\ref{fig:hype_properties}(c), and since $\lim_{q\to0}S_m(q)=\lim_{q\to0}S_1(\sqrt{m}\,q)=0$ whenever the subsets are hyperuniform, the hyperuniformity of a HyPE state is thus guaranteed by the hyperuniformity of its individual subsets.
This \textit{spectral self-similarity} reflects the preservation of long-range correlations under wavevector rescaling, analogous to scale-invariant behavior in critical phenomena.

In real-space, density fluctuations can be quantified by measuring the number variance within an observation window of size $R$.  Class I hyperuniform patterns~\cite{torquato2003} exhibit suppressed fluctuations characterized by the scaling relation $\sigma_N^2(R) = \Lambda R$, where $\Lambda$ is a surface-area constant that quantifies the extent of suppression of large-scale density fluctuations\cite{torquato2003,zachary2009}. $\Lambda$ is typically minimized in crystalline structures~\cite{torquato2003,zachary2009}.

The value of $\Lambda$, which increases with $p_0$ and remains independent of $K$, is shown in Fig.~S8.
For HyPE states, while the long-range behavior universally enforces $\sigma_N^2(R) \propto R$ scaling, the surface-area constant grows with $m$ as (Fig.~\ref{fig:hype_properties}(d)):
\begin{equation}
    \Lambda_m = \sqrt{m} \, \Lambda_1,
    \label{lambda_scaling}
\end{equation}
where $\Lambda_1$ is the surface constant of a single hyperuniform subset, and $\Lambda_m$ corresponds to that of the merged configuration formed by superimposing $m$ such subsets
\footnote{Equation~\eqref{lambda_scaling} assumes the number variance is measured after rescaling to unit number density. For HyPE states constructed from subsets with different number densities $\rho_s$ without rescaling them to unity, the surface constant obeys additivity and is given by
\[
\lambda_m \left(\sum_{s=1}^{m}  \rho_s\right) = \sum_{s=1}^{m} \lambda_s(\rho_s)
\]
where $\rho_s$ is the number density of subset $s$ and $\lambda_s$ and $\lambda_m$ are the surface constants measured without rescaling.}.
This construction enables continuous and unbounded control over $\Lambda_m$ via the parameter $m$, allowing one to tune the strength of hyperuniformity across a broad spectrum—from nearly crystalline (minimal $\Lambda$) to nearly Poissonian (large $\Lambda$)—without altering the long-range scaling law.
In this sense, HyPE provides a flexible framework for exploring hyperuniform states with tunable fluctuation amplitudes while preserving the defining asymptotic suppression of density fluctuations.

HyPE states fundamentally decouple local density (or area) fluctuations from long-range hyperuniformity. As a result, all short-range structural statistics in HyPEs converge to those of a PPP as $m$ increases.
In particular, the distribution of single-cell areas in HyPEs approaches a k-Gamma distribution of the form
\be
P(a) =  a^{k-1} e^{-a} /\Gamma(k),
\ee
with unit mean and shape parameter $k$ (Fig.~\ref{fig:hype_properties}(e). As $m$ increases, $k$ decreases and eventually saturates around $3.55$, yielding an area variance of $1/k \approx 0.28$ in the limit $m \to \infty$, which closely matches the analytically derived value for PPPs~\cite{brakke1987statistics}.
We further examined other short-range geometric descriptors of HyPEs and found similar convergence. The variance in cell topology (measured by the number of neighbors) approaches $1.78$, and the variance in cell perimeter converges to $0.94$, both in close agreement with PPP values~\cite{brakke1987statistics}. The distribution of nearest-neighbor distances $P(d_{\mathrm{nn}})$ likewise converges toward the PPP result as $m$ increases (Fig.~S12). Importantly, these results hold not only for HyPEs generated from EAR packings, but also for those constructed from biologically motivated model tissues (e.g., with $p_0 = 3.75$ and $K = 0.01$), which exhibit comparable values for area variance ($\sim 0.28$), topology variance ($\sim 1.78$), and perimeter variance ($\sim 0.93$) at $m = 300$.

\paragraph*{\textbf{ Entropic penalty associated with hyperuniformity } }
Beyond spatial correlations, the degree of structural disorder in a point pattern can be quantitatively assessed using information-theoretic measures. In particular, the Shannon entropy (see methods) provides a natural metric for the configurational randomness of a system. The PPP represents the maximum entropy state under fixed density constraints. In this context, hyperuniform states suppress long-wavelength fluctuations and are therefore expected to exhibit reduced configurational entropy. This raises a fundamental question: what is the entropic "cost" of achieving hyperuniformity?  To address this, we estimate the structural entropy of HyPE states using a method that infers entropy from the structure factor~\cite{ariel2020inferring}. Specifically, we compute the excess entropy per particle, $h$, defined relative to the entropy of the PPP (see Methods). Since the structure factor of HyPE states retains its functional form under the scaling $q \rightarrow q \sqrt{m}$, the entropy decreases systematically with increasing $m$, following a scaling relation $-h \propto m^{-1}$  (Inset of Fig.~\ref{fig:hype_properties}(c) and Fig.~S9, see derivation in methods). In the limit $m \rightarrow \infty$, $h$ vanishes, consistent with the maximal entropy of the PPP. These results quantify the minimal entropic penalty required for hyperuniformity, giving insight into the tradeoff between spatial order and configurational complexity in disordered systems.

\section*{Discussion}

Our findings reveal that hyperuniformity in confluent cellular systems emerges from a delicate competition between geometric frustration and mechanical stability, governed by two key parameters: cell cortical elasticity ($K$) and target cell shape ($p_0$). By tuning these variables, we engineer a continuum of hyperuniform states, spanning rigid solids to zero-shear-modulus fluids. This mechanical phase diagram
underscores hyperuniformity as a structural property that can be tuned independently of rigidity, with suppressed density fluctuations coexisting with tunable short-range disorder.
The mechanistic interplay between $K$ and $p_0$ highlights how biological systems might exploit hyperuniformity to optimize functionality. For instance, epithelial tissues near rigidity transitions ($p_0 \approx p_0^* $) could leverage   hyperuniformity to simultaneously achieve mechanical resilience and dynamic remodeling. The fluid hyperuniform phase mirrors the anomalous transport properties and density fluctuations observed in active glasses~\cite{Tjung_Berthier_PRL_2015,Hexner_Levine_PRL_2015}, suggesting a possible universal link between hyperuniformity and marginal stability in disordered systems.
A notable feature of the phase diagram is that at large $K$ hyperuniformity occurs on the fluid side of the transition rather than the solid side, even though the solid configurations have more evenly spaced cell centers. We stress that ``fluid'' here denotes a vanishing shear modulus obtained from linear response, and carries no dynamical connotation. The asymmetry is mechanical in origin: in the fluid, $\langle p_i \rangle$ tracks $p_0$, so $\delta p \ll \sigma_p$ and the perimeter term exerts little net tension, allowing the area constraint to dominate and suppress $\sigma_a$ (Figs.~S5,~S6). Hyperuniformity can even be achieved in the deep fluid regime, where cells may become arbitrarily close to one another. Low area fluctuations only prevent cells from being simultaneously close to all of their Voronoi neighbors, while allowing them to be arbitrarily close to some, as long as they maintain greater distances from others.

Given that low area fluctuations underlie hyperuniformity, it is crucial to reduce cell size heterogeneity. Cell size homeostasis is tightly regulated, with both mass-dependent cell cycle control and growth modulation helping to minimize size variation~\cite{liu2024cell}. A recent experiment in the fruit fly wing epithelium demonstrates that, following the arrest of cell divisions, cellular packing becomes increasingly uniform, with cell areas progressively concentrating around the mean~\cite{chhajed2025cell}. Moreover, shear flow promotes size uniformity, as its reduction increases cell size polydispersity~\cite{chhajed2025cell}.
Osmotic pressure and substrate stiffness also influence cell volume and its fluctuation~\cite{guo2017cell}.

Our simulations are based on energy minimization, excluding the effects of temperature and active noise. However, the structure factor in real systems should be decomposed into a configurational part and a fluctuating part~\cite{ikeda2015}. The configurational part, an “inherent” property of the amorphous packing~\cite{stillinger1982hidden}, is associated with the averaged position of the cell center and independent of temperature. In contrast, the fluctuating part captures deviations from the average positions and exhibits a strong temperature dependence, vanishing at zero temperature. Therefore, reducing thermal fluctuations or isolating the configurational component could facilitate the observation of hyperuniformity in natural cellular systems. Notably, the work ~\cite{ikeda2015} also suggests that hyperuniformity is unrelated to jamming criticality, and our results further indicate that hyperuniformity is independent of the solid–fluid transition.
Each configuration is a local minimum, and the statistics we report are ensemble averages over minima reached from independent random initial conditions (Fig.~S2), rather than properties of a single basin.

Also central to this work is the introduction of Hyperuniform Poisson Ensemble (HyPE) states, which preserve long-range uniformity while asymptotically approaching Poissonian randomness at short scales. These states exhibit a surface-area constant ($\Lambda$) that scales universally with the number of superimposed subsets ($m$), enabling control over fluctuation amplitudes without altering hyperuniformity. HyPEs minimize the entropic cost of hyperuniformity while maximizing structural randomness, revealing a thermodynamic trade-off inherent to disordered hyperuniform systems. This makes HyPEs a lower bound for hyperuniformity, akin to an "ideal gas" of hidden-order materials.

Future work should explore dynamical extensions of this model, such as incorporating activity through a finite self-propulsion speed $v_0>0$, to test whether hyperuniformity in tissue-like materials emerges as a nonequilibrium steady state. We emphasize that every configuration reported in this work is an athermal, force-balanced energy minimum; the term ``nonequilibrium'' refers throughout to this prospective extension, and not to the states studied here.
This is not merely a robustness check: noise and activity can qualitatively alter hyperuniform behavior in other settings~\cite{deluca2024,lei2024}, so the classification obtained here for athermal minima need not carry over unchanged to a noisy or self-propelled tissue, and the response of the low-$q$ plateau to noise is a natural place to look.
Experimentally, colloidal systems with tunable interactions could validate the phase diagram, while live-tissue imaging might probe hyperuniformity in developing embryos. Finally, the entropy scaling of HyPEs invites theoretical inquiries into information encoding in disordered systems, linking Shannon entropy to material functionality.

Our framework bridges cellular mechanics and condensed matter physics to guide multifunctional material design.
Hyperuniformity alone does not guarantee a photonic band gap: the short-range structure and the class of hyperuniformity both matter~\cite{klatt2022,li2018a}. Because $m$ tunes the short-range order of HyPE states continuously while leaving the long-range hyperuniformity intact, it offers a route to varying photonic response independently of the large-scale density fluctuations.
Solid hyperuniform states, with geometric frustration, may yield auxetic or shape-memory composites under strain. In biology, hyperuniform fluid phases in epithelia suggest a regulatory mechanism for tissue homeostasis—suppressed fluctuations may stabilize dynamics, a process potentially disrupted in pathologies like cancer or fibrosis.

\section*{Acknowledgments}
We acknowledge support from the National Science Foundation (Grant Nos. DMR-2046683 and PHY-2019745), the National Institute of General Medical Sciences of the National Institutes of Health (Grant No. R35GM150494), the Alfred P. Sloan Foundation, and the Human Frontier Science Program (Reference No. RGP0007/2022). This research was supported in part by grant NSF PHY-2309135 to the Kavli Institute for Theoretical Physics (KITP).

\section*{Methods}
\subsection*{\label{app:simu} Simulating a confluent tissue monolayer}

To simulate the confluent tissue, $N$ cells are randomly placed within a square box of size $L =\sqrt{N}$ under periodic boundary conditions. The average area per cell is unity. The default value for $N$ is 2430 unless otherwise specified.
All simulations are performed at zero temperature and zero self-propulsion ($v_0=0$); no thermal or active noise enters at any stage, so the configurations reported here are athermal, force-balanced energy minima rather than members of a finite-temperature or nonequilibrium steady-state ensemble.
Through the open-source software cellGPU~\cite{sussman2017}, the fast Inertial Relaxation Engine(FIRE) algorithm~\cite{guenoleAssessmentOptimizationFast2020} is utilized to minimize the energy and obtain disordered ground states of the SPV model.  We explore a range of cortical elasticity $K$ values spanning from $0$ to $1$ and target shape index $p_0$ from $3.6$ (solid) to $4.6$ (fluid). A $p_0$ value of 4.6 corresponds to the perimeter of an equilateral triangle with unit area. In both cases, with $K=0$ and the solid side of the transition, all structures undergo thorough minimization, ensuring that the maximum residual forces remain below $10^{-13}$. On the fluid side of the transition the energy landscape is nearly flat and minimization converges more slowly, increasingly so at larger $K$. There we terminate when the maximum residual force falls below $10^{-5}$, or at the iteration limit of $10^6$, whichever is reached first. This threshold is five orders of magnitude below the natural force scale of Eq.~\eqref{eq:epsilon}, which is of order unity in these units. We further note that the low-$q$ plateau of $S(q)$ in the fluid varies systematically with $K$ over several decades (Fig.~\ref{fig:sq_regimes}(b)), a dependence that a fixed numerical tolerance could not produce.
Additionally, we examine the independence on the system size in Fig.~S1 and the initial condition in Fig.~S2.

\subsection*{\label{app:hu}Quantifying Hyperuniformity}
\paragraph{Structure Factor}
The structure factor $S(\vec{q})$ is given by\cite{hansen2013chapter},
\begin{equation}
    S(\vec{q})=\frac{1}{N}\left\langle\sum_{j=1}^{N} \sum_{k=1}^{N} \mathrm{e}^{-i \vec{q}\left(\vec{r}_{j}-\vec{r}_{k}\right)}\right\rangle
\end{equation}
where $\vec{r}_{j}$ is the position of $j^{th}$ particle, and $\left\langle \cdot \right\rangle$ represents the ensemble average.
$S(q)$ is the spherical average of $S(\vec{q})$ within reciprocal space at a radius $q$. The vanishing structure factor in the zero wave number limit, $\lim_{q\rightarrow0}S(q)=0 $ defines a system as hyperuniform~\cite{torquato2003}.

\subsection*{Effective Hyperuniformity Index}

It is a challenge to attain a rigorous statistical evaluation for hyperuniformity, largely due to the definition of the structure factor $S(q)$ at infinite wavelength. Yet both numerical simulations and lab experiments encounter limitations from system size constraints and errors or noise inherent in measurements. To quantify these non-idealized samples how close to hyperuniformity, a hyperuniformity index $H$~\cite{torquato2018} is raised,
\begin{equation}
    H\equiv\frac{S(q\rightarrow 0)}{S(q_{peak})},
\end{equation}
where $S(q\rightarrow 0)$ is the estimated value of the structure factor at the origin and $S(q_{peak})$ is the value of the structure factor at the first dominant peak.
A threshold is necessary because in any finite system $S(q)$ is sampled only at discrete wavevectors $q \ge 2\pi/L$ and never vanishes identically; hyperuniformity inferred from finite data is therefore necessarily an operational statement. The index $H$ provides a dimensionless, normalization-independent measure of how strongly long-wavelength fluctuations are suppressed relative to the dominant structural length scale.
 The empirical condition $H < 10^{-4}$ is widely adopted as the criterion for "effective hyperuniformity" in finite systems\cite{atkinson2016}, which we employ here.
This criterion originates in Ref.~\cite{atkinson2016} and is stated in the form used here in Ref.~\cite{torquato2018}; both note explicitly that the numerical value of the threshold is a convention rather than a sharp criterion. The same value of $10^{-4}$ is adopted in subsequent work~\cite{kim2018,torquato2021}.
Crucially, once non-strict hyperuniformity is permitted, the specific numerical value of this $H$ threshold does not affect our claim: for the representative states of Fig.~S1 the measured $H$ and $H'$ lie one to two decades below $10^{-4}$, so shifting the threshold by an order of magnitude would not reclassify them.

In the Voronoi model, the structure factor displays a plateau at small wavenumbers in the large-scale limit, as clearly seen on a logarithmic scale. This plateau reflects a subtle but statistically significant nonmonotonic behavior, preventing the system ($K>0$) from being strictly hyperuniform. Yet it still allows for characterization using an effective hyperuniformity measure.
Because of this nonmonotonicity, the wavevector at which $S(q)$ attains its minimum over the sampled range does not always coincide with the smallest wavevector resolved by the simulation box, $q_{\min}=2\pi/L$. We therefore report two related estimators of $H$: $H\equiv \min_{q} S(q)/S(q_{peak})$, built from the global minimum of $S(q)$ over the sampled range, and $H'\equiv S(q_{\min})/S(q_{peak})$, built from the value of $S(q)$ evaluated strictly at $q_{\min}=2\pi/L$. Comparing $H$ and $H'$ further indicates where the suppression of $S(q)$ occurs: $H \approx H'$ signals the minimum of $S(q)$ happens at $q_{\min}$, consistent with genuine long-wavelength suppression, whereas $H \ll H'$ indicates effective hyperuniformity.
We denote this wavevector $q^{*}=\argmin_q S(q)$. Its dependence on system size is shown in the left insets of Fig.~S1. Where $S(q)$ turns up below the plateau, $q^{*}$ separates from $q_{\min}$ and becomes independent of $N$ once the system is large enough to resolve it, locating the onset of the plateau. Where the plateau is flat, $q^{*}$ instead tracks $q_{\min}$ at all $N$, as it does for the EAR packings; in that case the presence of a plateau is established not by $q^{*}$ but by the saturation of $H$ and $H'$.
The $N$ dependence of $H$ and $H'$ is shown in the right insets of Fig.~S1: neither grows with system size. For EAR packings both continue to decrease, as expected for strict hyperuniformity, whereas for $K>0$ they approach finite values once the box is large enough to resolve the plateau, so the classification does not depend on the system size at which it is evaluated.

\section*{\label{app:GB}Elastic moduli}

At the tissue level, its mechanical response is characterized by the shear modulus $G$. A non-zero $G$ corresponds to a solid-like tissue while $G$ vanishes for a fluid state.   We obtain $G$ by calculating the  linear response  to an infinitesimal affine strain $\gamma$ via the Born-Huang formulation~\cite{Maloney_PRE_2006,huang2022,nguyen2025},

\begin{eqnarray}
  G = G_{\text{affine}}-G_{\text{non-affine}} \nonumber
  = \frac{1}{A_{total}} \left[\frac{\partial^{2}E}{\partial \gamma^{2}}  - \Xi_{i\mu}M^{-1}_{i\mu j\nu}\Xi_{j \nu} \right]_{\gamma=0}.
  \label{eq:sm}
\end{eqnarray}
In Eq.~\eqref{eq:sm},

$\Xi_{i\mu}$ is the derivative of the  force on vertex $i$ with respect to strain given by
\be
\Xi_{i\mu} \equiv \frac{\partial^{2}E}{\partial \gamma \partial r_{i\mu}},
\ee
where $ r_{i\mu}$ is the position of vertex $i$ and $\mu=x,y$ is the Cartesian index.
$A_{total}=\sum_i^N A_i$ is the total area of the tissue.
$M$ is the Hessian matrix given by the second derivative of the tissue energy $E$ with respect to position vectors of vertices $i$ and $j$
\be
M_{i\mu j\nu}=\frac{\partial^{2}E}{\partial r_{i\mu} \partial r_{j\nu}}.
\ee
The same Hessian determines the linear relaxation of these configurations. For overdamped dynamics the relaxation rates about a force-balanced minimum are the eigenvalues of $M$ divided by the friction coefficient, so the modes whose eigenvalues vanish as $p_0 \to p_0^*$ are both the origin of the vanishing shear modulus and the slowest to relax.
To calculate the bulk modulus, we consider the isotropic compression process, $x \to x' = \lambda x, \; y \to y' = \lambda y$. Here $\lambda$ is compressional ratio along $x$ and $y$ directions.
So the strain tensor is defined by
\be
\begin{pmatrix}
\gamma_{xx} & \gamma_{xy} \\
\gamma_{yx} & \gamma_{yy}
\end{pmatrix}
=
\begin{pmatrix}
\lambda -1 & 0 \\
0 & \lambda - 1
\end{pmatrix}
\ee

Similar to the shear modulus $G$, we also calculate the bulk modulus using the linear response to an infinitesimal strain tensor.

\begin{eqnarray}
B = B_{\text{affine}}-B_{\text{non-affine}} \nonumber
= \frac{1}{2 A_{total}}\left\{ \left[\frac{\partial^{2}E}{\partial \gamma_{xx}^{2}}  - \Xi_{i\mu}M^{-1}_{i\mu j\nu}\Xi_{j \nu}\right]_{\gamma_{xx}=0} \right. \nonumber
 \left. + \left[\frac{\partial^{2}E}{\partial \gamma_{yy}^{2}}  - \Xi_{i\mu}M^{-1}_{i\mu j\nu}\Xi_{j \nu}\right]_{\gamma_{yy}=0} \right\}
  \label{eq:bm}
\end{eqnarray}

Because $\gamma_{xx} = \lambda - 1$, the derivation of $E$ with respect to $\gamma_{xx}$ would be the same with that of $\lambda$ at the limit $\lambda \to 1$. Finally, we have the bulk modulus
\begin{eqnarray}
B = B_{\text{affine}}-B_{\text{non-affine}} \nonumber
= \frac{1}{2 A_{total}}\left\{ \left[\frac{\partial^{2}E}{\partial \lambda^{2}}  - \Xi_{i\mu}M^{-1}_{i\mu j\nu}\Xi_{j \nu}\right]_{\lambda=1} \right. \nonumber
 \left. + \left[\frac{\partial^{2}E}{\partial \eta^{2}}  - \Xi_{i\mu}M^{-1}_{i\mu j\nu}\Xi_{j \nu}\right]_{\eta=1} \right\}
  \label{eq:bm1}
\end{eqnarray}

\section*{\label{app:SRO} The Pair Correlation}
The pair correlation function is~\cite{hansen2013chapter2}
\begin{equation}
    g(r)=\frac{1}{N\rho}
    \langle
    \sum_j\sum_{j\neq k}\delta (\vec{r}+\vec{r_j}-\vec{r_k})
    \rangle
\end{equation}
where $\rho$ is the average particle density of the isotropic system, $\vec{r}_{j}$ is the position of $j^{th}$ particle, and $\left\langle \cdot \right\rangle$ represents the emsemble average.

\section*{\label{app:entropy} Structural Entropy}
The excess entropy per particle relative to the ideal gas entropy~\cite{ariel2020inferring}, is given by $h \equiv (\mathcal{H} - \mathcal{H}_{\text{ideal}}) / N_t$, where $\mathcal{H}$ is the Shannon entropy and $N_t$ is the total number of particles. The excess entropy is directly related to the structure factor $S(q)$\cite{ariel2020inferring}:
\begin{equation}
h[S(q)] = \frac{1}{2 (2\pi)^d \bar{\rho}} \int d \vec{q} \left( \ln S(q) - S(q) + 1 \right)
\end{equation}
where $d$ is the spatial dimensionality, and the mean density $\bar{\rho} = 1$.
For the integration, we use $dq = 1/\sqrt{N_t}$.
The calculated values are then shifted to align the excess entropy $h$ for PPP across all sizes with the ideal gas limit, setting $h = 0$. This adjustment effectively eliminates errors related to size effects during the integration process.
Notably, the mathematical convergence of the excess entropy integral is maintained under strict hyperuniformity.
Although $\mathbf{\ln S(q)}$ asymptotically approaches $\mathbf{-\infty}$ near the origin (i.e., $\mathbf{\ln S(q) \sim \ln q^\alpha}$ for $q \to 0$), its integral near the lower boundary $\mathbf{(q \to 0)}$ converges to zero.

Recalling the scaling relation for the structure factor of HyPE states (Eq.\eqref{m-scaling}), we can derive a corresponding scaling relation for the entropy in $d$-dimensions
\begin{eqnarray}
h[S_m(q)] &=& \frac{1}{2 (2\pi)^d \bar{\rho}} \int d^d \vec{q} \left( \ln S_m(q) - S_m(q) + 1 \right)\\
&=& \frac{1}{2 (2\pi)^d \bar{\rho}} \int d^d \vec{q} \left( \ln S_1(q \sqrt{m}) - S_1(q \sqrt{m}) + 1 \right) \\
&=& \frac{1}{2 (2\pi)^d \bar{\rho}} \int \frac{d^d \vec{q}}{m^{d/2}} \left( \ln S_1(q) - S_1(q) + 1 \right)
=m^{-d/2} h[S_1(q)].
\end{eqnarray}

\section*{Data Availability}
The data supporting the findings of this study, comprising the cell-center configurations generated by the model together with the structure factors derived from them, are available at Zenodo under DOI 10.5281/zenodo.21937906. Source data are provided with this paper.

\section*{Code Availability}
The vertex model simulations were performed with the open-source software cellGPU~\cite{sussman2017}. The custom code used to compute the structure factors, their circular averages and the pair correlation functions is deposited together with the data at Zenodo under DOI 10.5281/zenodo.21937906.
\clearpage
\bibliography{nat_comm_revision_Aug14}

\clearpage
\section*{Supplemental Information}
\setcounter{figure}{0}
\renewcommand{\figurename}{Fig.}
\renewcommand{\thefigure}{S\arabic{figure}}
\setcounter{figure}{0}
\renewcommand{\figurename}{Fig.}
\renewcommand{\thefigure}{S\arabic{figure}}
\begin{figure}[ph]
    \centering
    \includegraphics[width=\columnwidth]{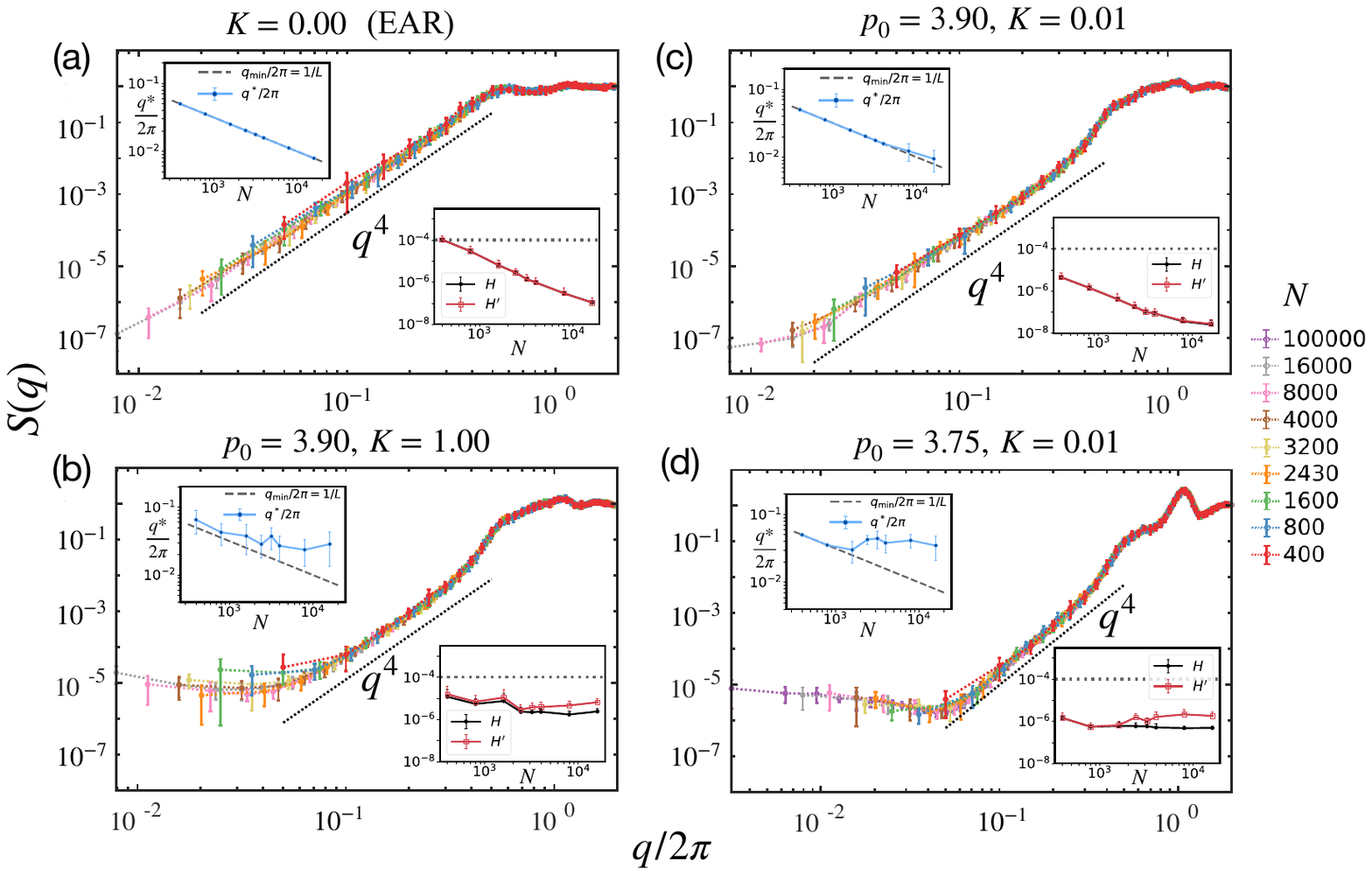}
\caption{
\textbf{Finite-size scaling of the structure factor in representative states.}
Each panel shows $S(q)$ for system sizes $N = 400$ to $16000$, for
(a) EAR packings ($K=0$), (b) $p_0=3.90$, $K=1.00$, (c) $p_0=3.90$, $K=0.01$, and
(d) $p_0=3.75$, $K=0.01$ ($N = 400$ to $16000$). Left insets: the wavevector $q^{*}=\argmin_q S(q)$ as a
function of $N$, together with the smallest wavevector accessible at that system size,
$q_{\min}=2\pi/L$. Right insets: the $N$ dependence of
$H \equiv \min_{q} S(q)/S(q_{peak})$ and $H' \equiv S(q_{\min})/S(q_{peak})$; the dashed
line marks the $10^{-4}$ threshold for effective hyperuniformity. Error bars denote the
standard deviation over independent realizations.
For EAR packings $q^{*}$ coincides with $q_{\min}$ at every $N$ and both indices decrease
without saturating, as expected for strict hyperuniformity with $S(q)\sim q^{4}$. For
$K>0$, $q^{*}$ separates from $q_{\min}$ and becomes independent of $N$ once the box is
large enough to resolve it, and $H$ and $H'$ saturate at finite values, identifying a
plateau of finite height and locating the wavevector below which it sets in.
}
    \label{fig:finite_size_scaling}
\end{figure}
\clearpage

\begin{figure}[p]
    \centering
    \includegraphics[width=.85\columnwidth]{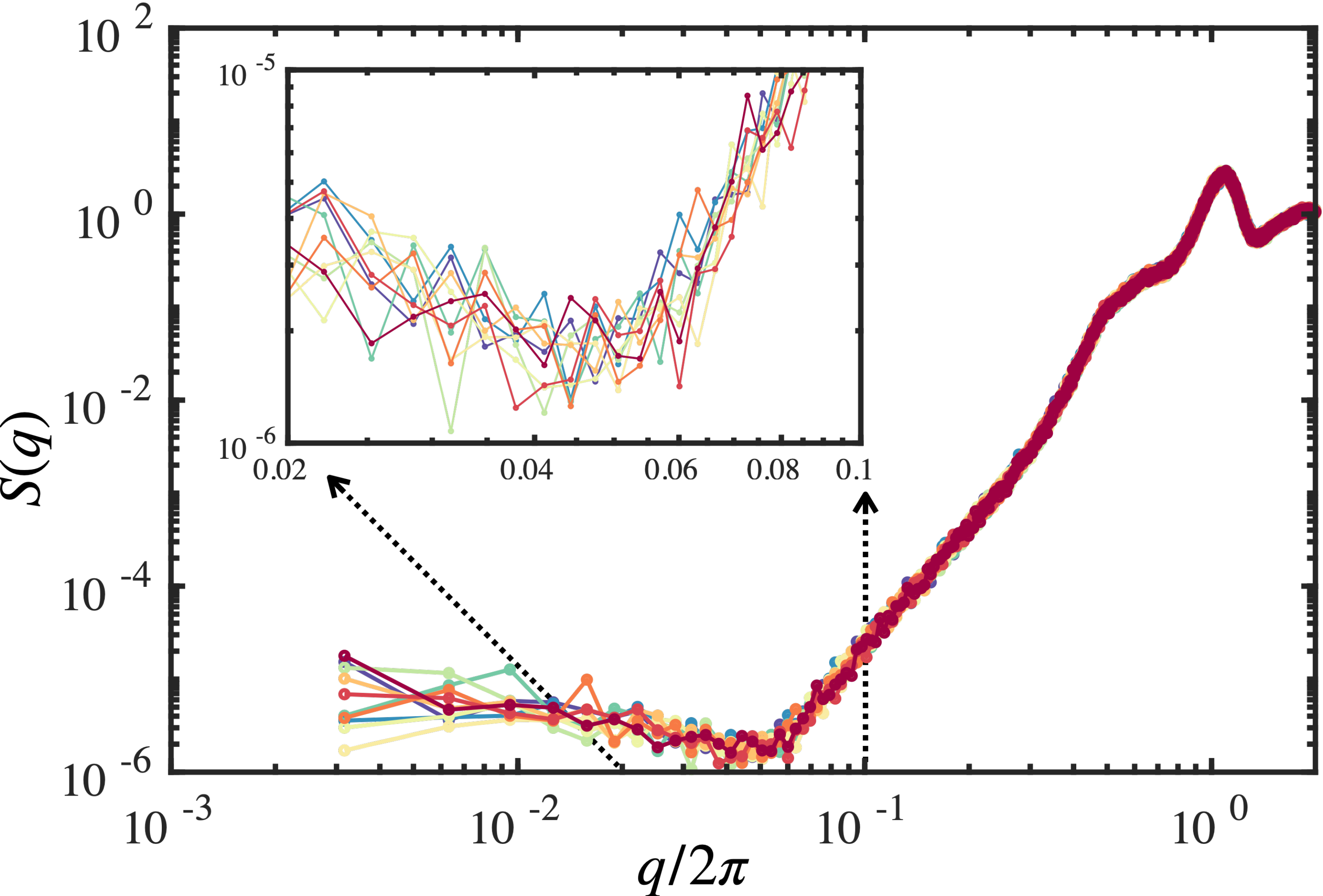}
    \caption{\textbf{The robustness of the structure factor across various random initial configurations.} }
    \label{fig:s_q_seeds}
\end{figure}
\clearpage

\begin{figure}[p]
    \centering
    \includegraphics[width=\columnwidth]{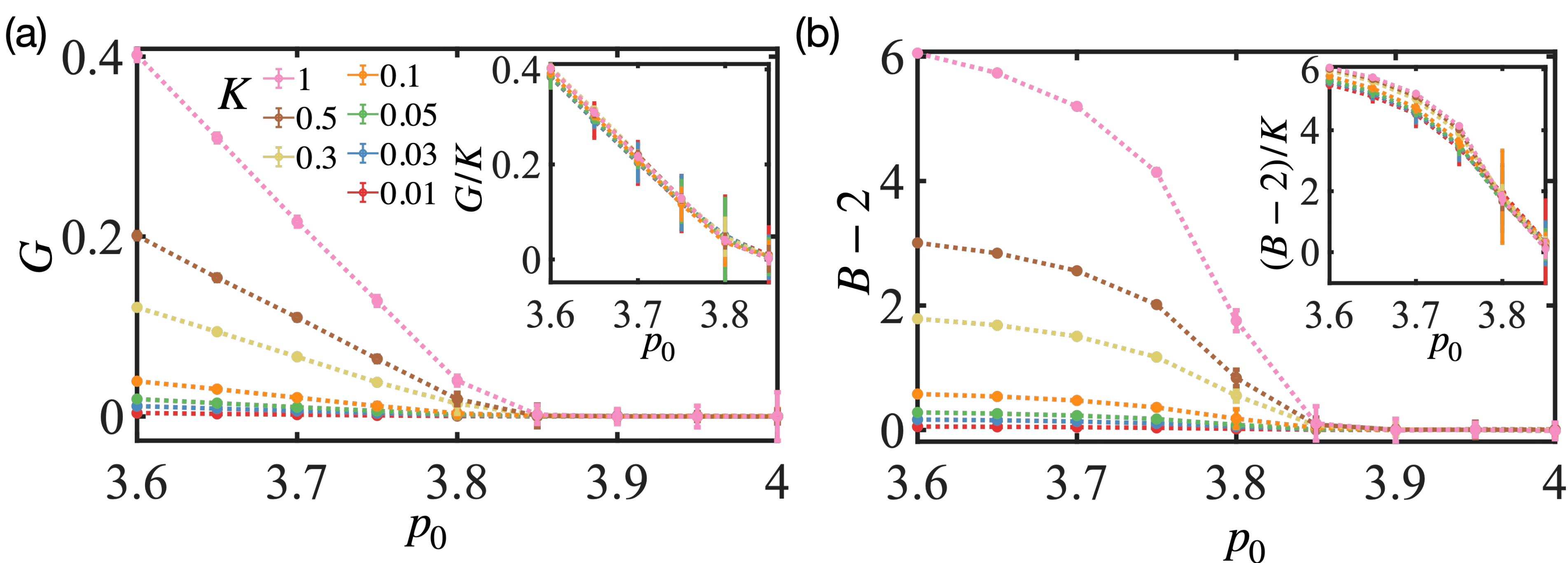}
    \caption{\textbf{Elastic moduli characterize the tissue's rigidity.}
    (a) Shear modulus $G$ and (b) bulk modulus $B-2$ versus $p_0$ for different $K$. Insets: the rescaled quantities $G/K$ (inset of (a)) and $(B-2)/K$ (inset of (b)), showing collapse onto a single polyline.
    As $p_0$ increases, the tissues become less solid. Both the shear modulus $G$ and the bulk modulus $B$ decrease, with the rate of increase  scaling linearly with $K$. Ultimately, the tissues transit into the fluid phase, featured by vanished shear modulus $G=0$, and constant bulk modulus $B=2$. }
    \label{fig:elastic_moduli}
\end{figure}
\clearpage

\begin{figure}[p]
    \centering
    \includegraphics[width=\columnwidth]{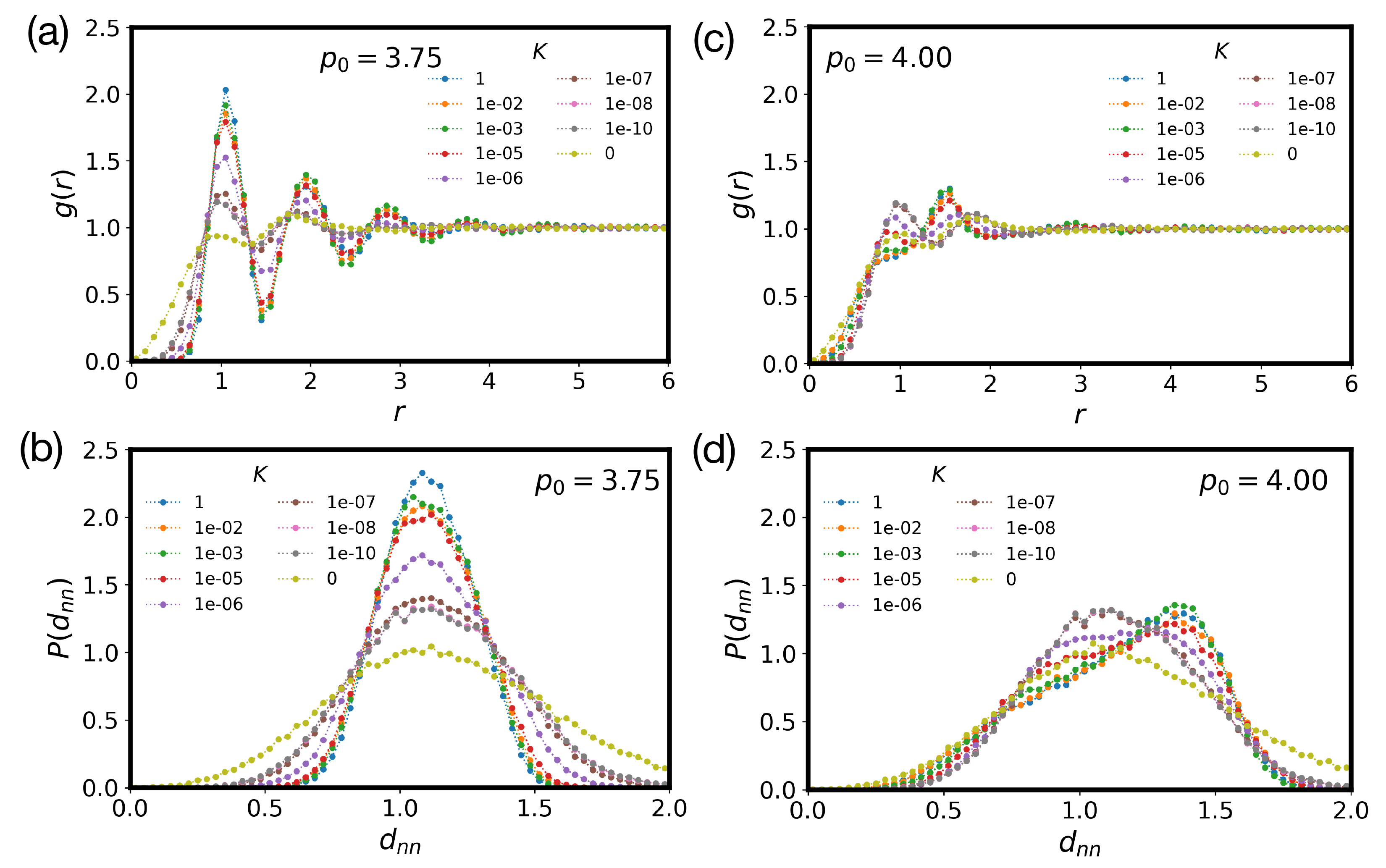}
    \caption{\textbf{Cortical elasticity effect on the short-range order.} In both (a,b) the solid phase ($p_0=3.75$) and (c,d) the fluid phase ($p_0=4.00$), (a,c) the pair correlation function $g(r)$ and (b,d) the probability density function of the distance between neighboring cells $d_{\mathrm{nn}}$ are near identical for $K \gtrsim 10^{-5}$ and vary smoothly for smaller $K$, approaching but remaining distinguishable from the $K=0$ (EAR) result.}
    \label{fig:sro_vs_K}
\end{figure}
\clearpage

\begin{figure}[p]
    \centering
    \includegraphics[width=.85\columnwidth]{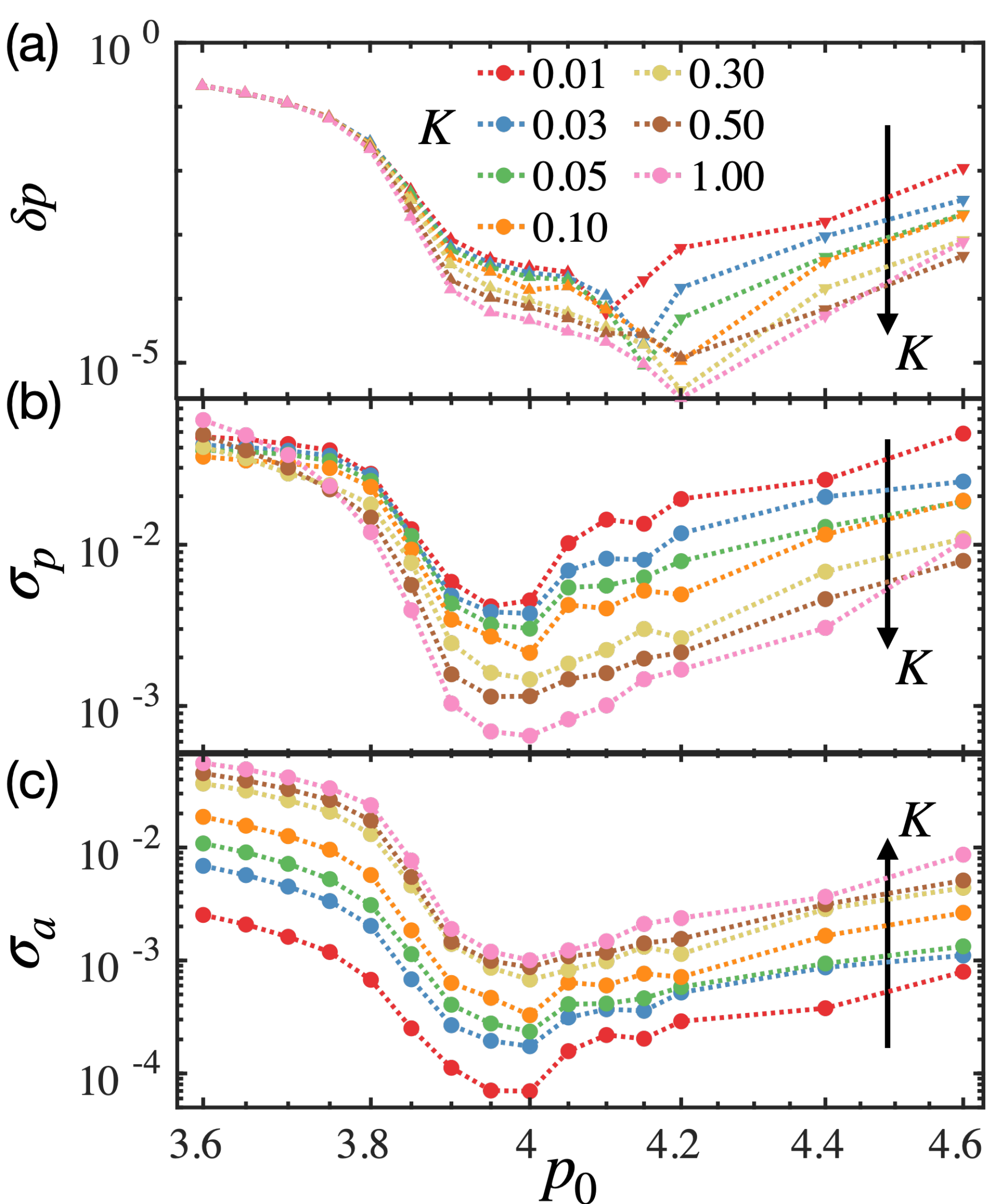}
    \caption{\textbf{The distribution of cell perimeters and areas.}
    (a) The deviation from the target shape index, denoted as $\delta p = \langle p_i - p_0 \rangle$, (b) the standard deviation of cell perimeters $p_i$, and (c) the standard deviation of cell areas $a_i$, are examined across various values of $K$ as functions of the target shape index $p_0$.
    }
    \label{fig:shape_distributions}
\end{figure}
\clearpage
 
\begin{figure}[p]
    \centering
    \includegraphics[width=.85\columnwidth]{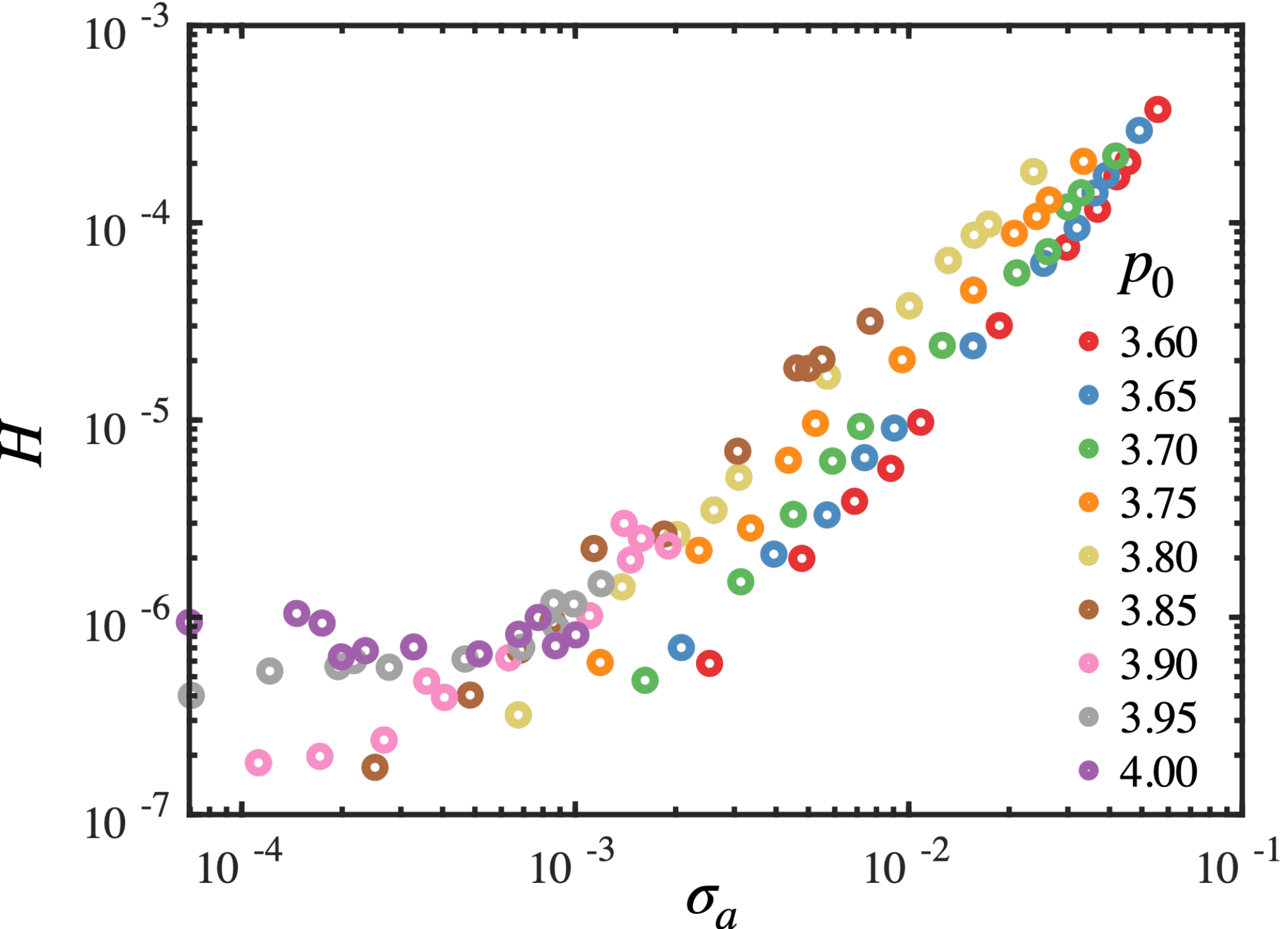}
    \caption{\textbf{Area fluctuation $\sigma_a$ effectively characterizes the hyperuniformity.} $\sigma_a$ presents a positive correlation with hyperuniformity index $H$ across various $p_0$. When $\sigma_a>10^{-3}$, $H$ increases with increased $\sigma_a$.}
    \label{fig:sigma_a_vs_H}
\end{figure}
\clearpage

\begin{figure}[p]
    \centering
    \includegraphics[width=.85\columnwidth]{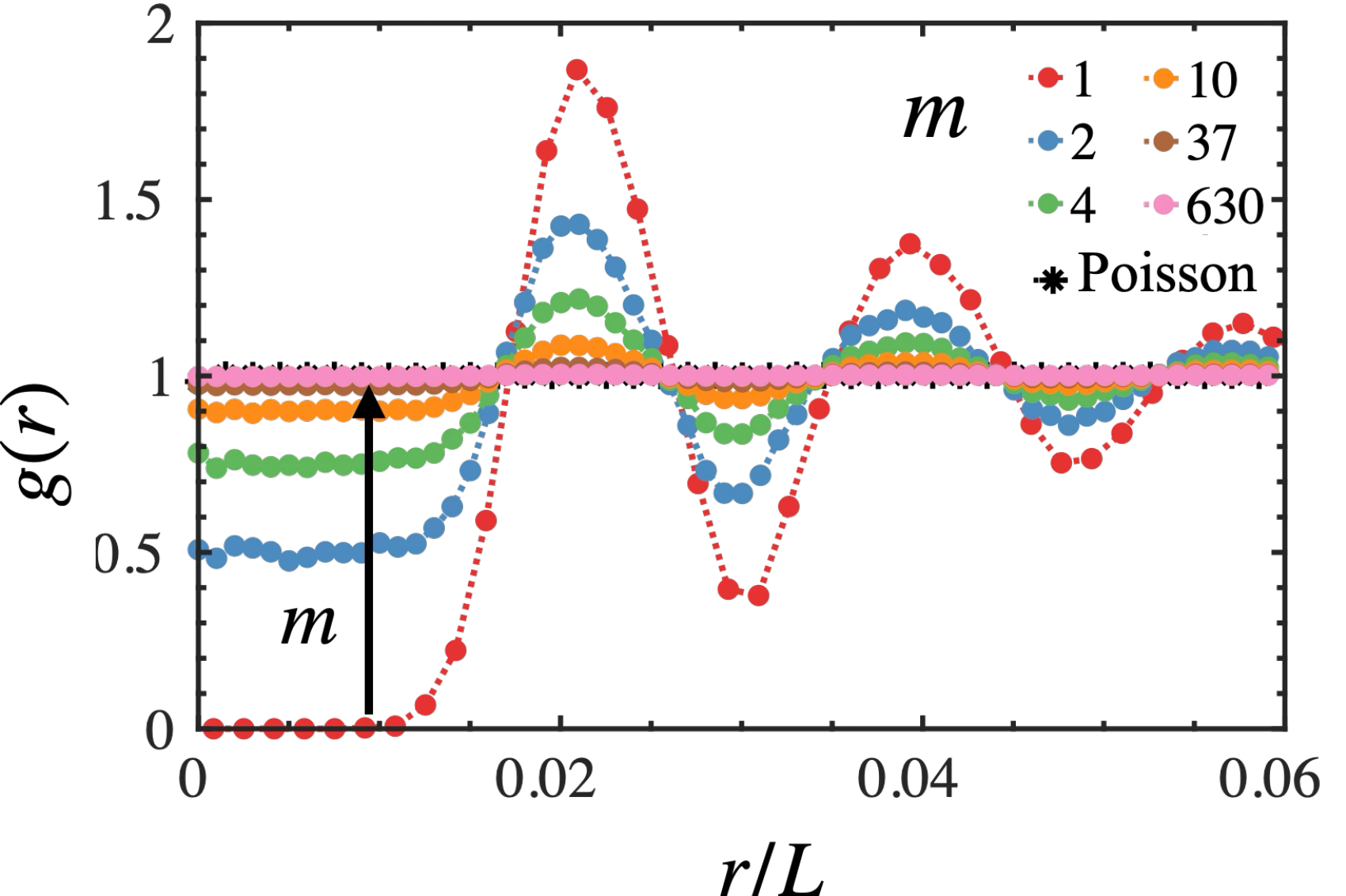}
    \caption{\textbf{The pair correlation function $g(r)$ for HypE states at various numbers of superimposed point sets.} The HyPE state is constructed by merging $m$ hyperuniform subsets, each characterized by parameters $p_0 = 3.75$ and $K = 0.01$.}
    \label{fig:hype_gr}
\end{figure}

\clearpage

\begin{figure}[p]
    \centering
    \includegraphics[width=.85\columnwidth]{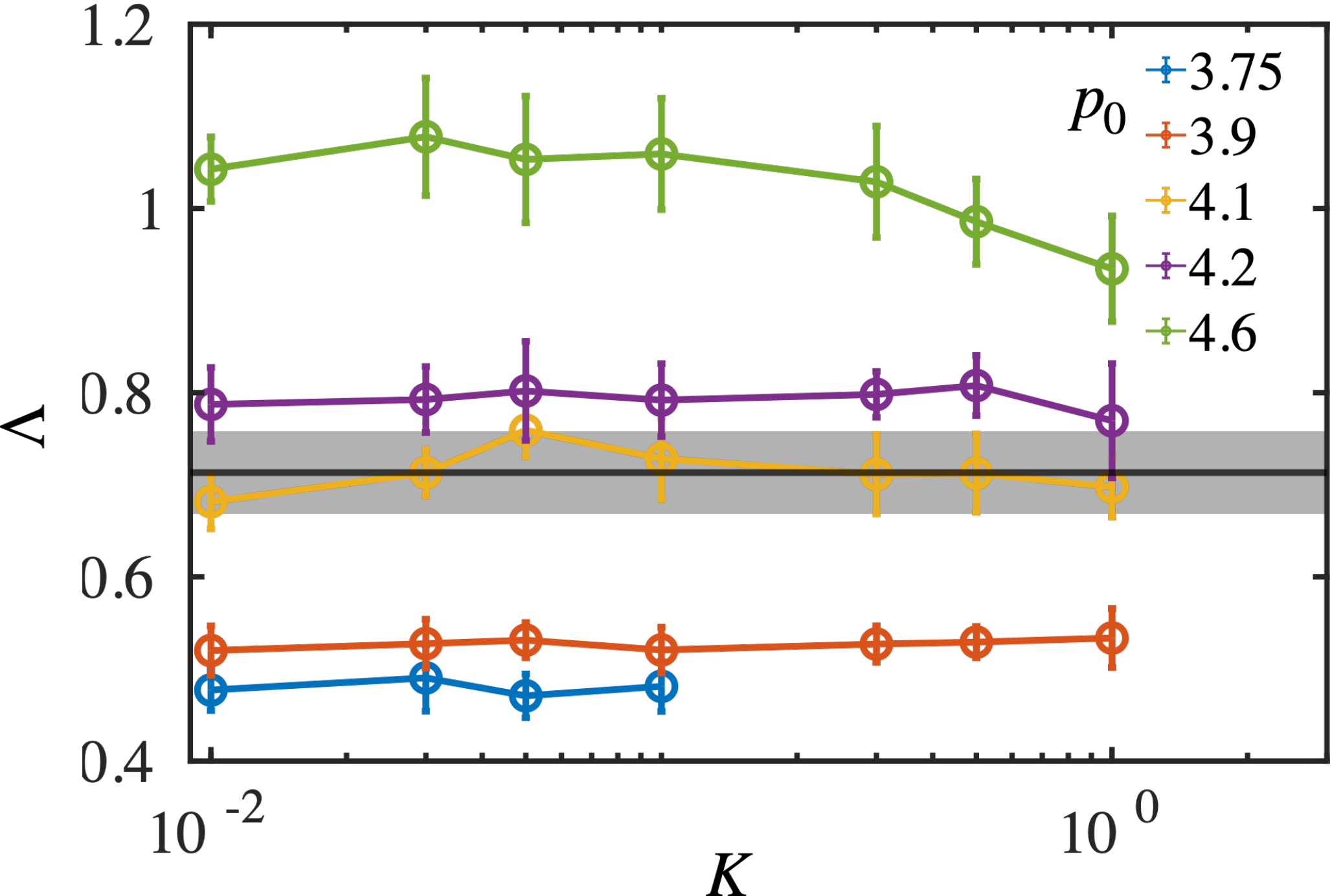}
    \caption{\textbf{Surface-area Constant $\Lambda$} for hyperuniform states derived directly from the Voronoi model. $\Lambda$ increases with $p_0$ increases but is insensitive to $K$. The black line with a gray shadow represents the surface constant $\Lambda$ for EAR packings.
    }
    \label{fig:surface_constant}
\end{figure}
\clearpage

\begin{figure}[p]
    \centering
    \includegraphics[width=0.85\columnwidth]{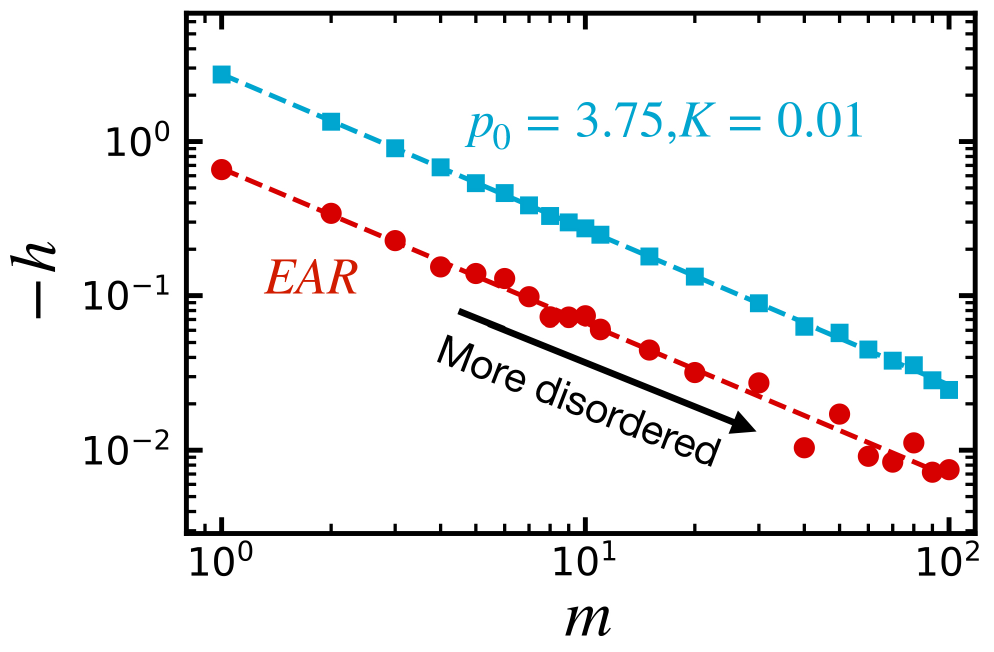}
    \caption{\textbf{Hidden Information of HyPE states.}
  The excess entropy per cell relative to the ideal gas entropy $h$ is shown as a function of the number $m$ of merged subsets used to construct each HyPE: EAR-packing configurations (red circles) and hyperuniform subsets with $p_0 = 3.75$ and $K = 0.01$ (blue squares).  Each subset comprises $2430$ cells. The fitting results suggest a scaling behavior of the form $-h \propto m^{-1}$.
    }
    \label{fig:hype_entropy}
\end{figure}
\clearpage

\begin{figure}[p]
    \centering
    \includegraphics[width=0.9\columnwidth]{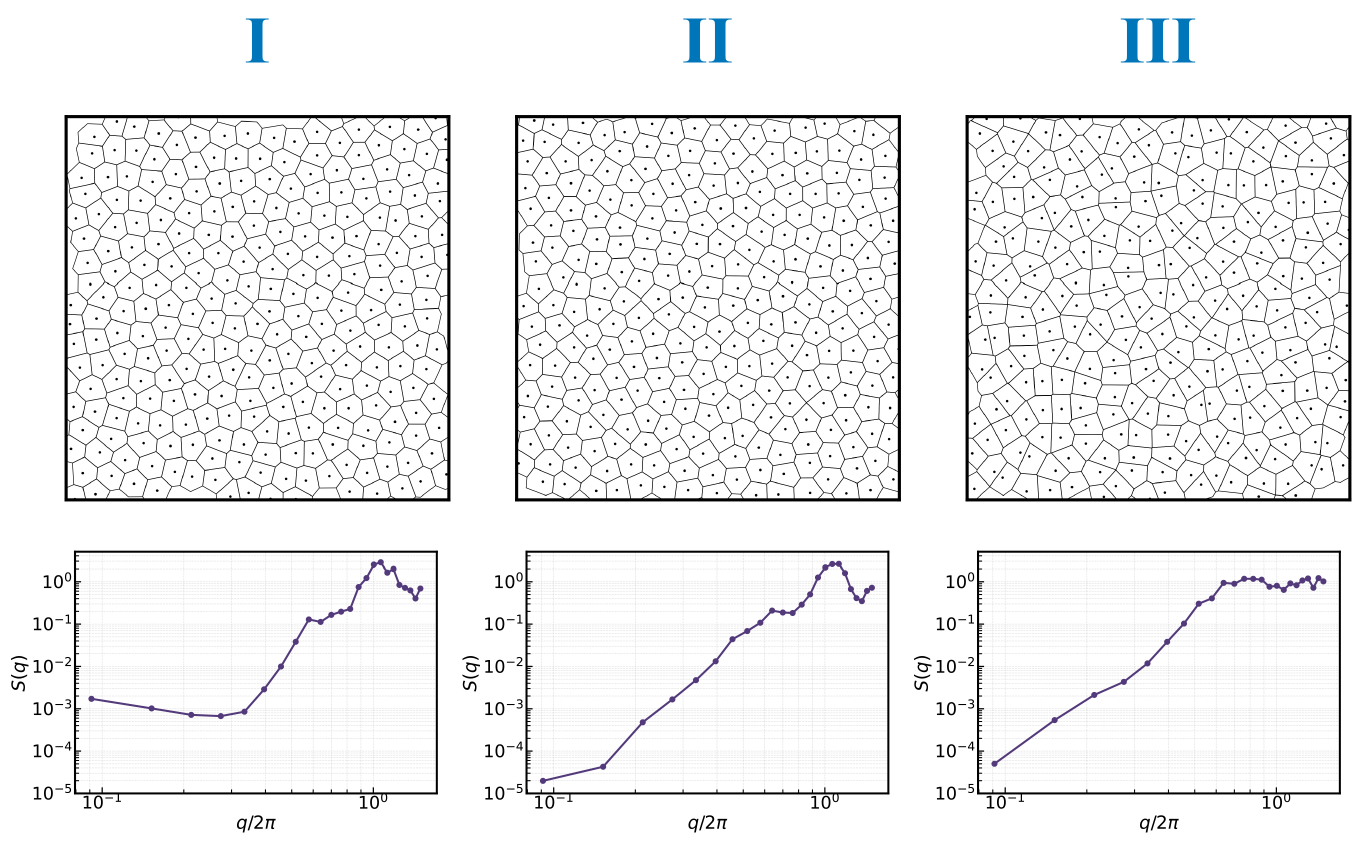}
    \caption{\textbf{Representative Samples of Phase Diagram.}
    Three samples from the phase diagram Fig.~4: (I) non-hyperuniform solid state, generated with $p_0=3.7$ and $K = 0.8$, (II) hyperuniform solid state, generated with $p_0=3.7$ and $K = 0.03$, and (III) hyperuniform fluid state generated with $p_0=4$ and $K = 0.1$.
    Top row: real-space Voronoi tessellations, with dots marking cell centers. Bottom row: the corresponding angularly averaged structure factor $S(q)$. States I and II share the same $p_0$ and differ only in $K$; the difference between them is not visible in the local cell geometry but appears at small $q$, where $S(q)$ plateaus near $10^{-3}$ in I while continuing to decrease below $10^{-5}$ in II.
    }
    \label{fig:representative_configs}
\end{figure}
\clearpage

\begin{figure}[p]
    \centering
    \includegraphics[width=1\columnwidth]{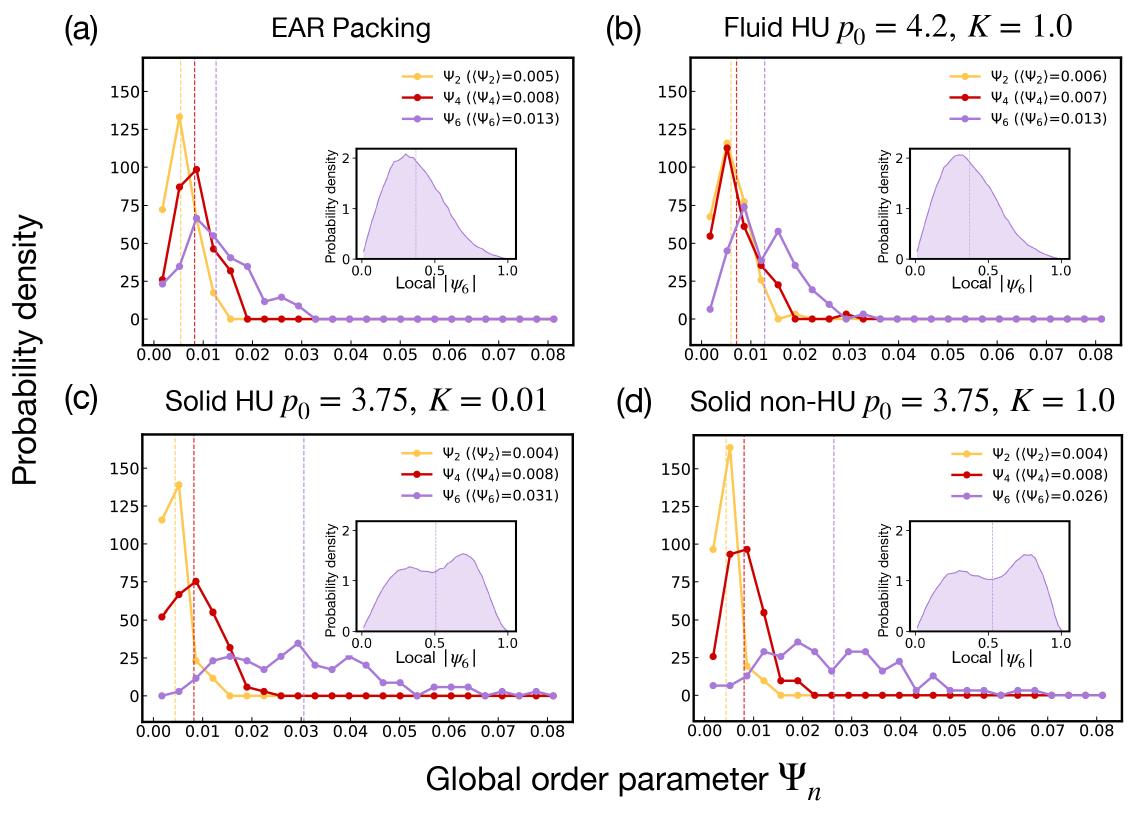}
    \caption{\textbf{Distribution of orientational order parameters.}
    Distribution of global orientational order parameters $\Psi_2, \Psi_4,$ and $\Psi_6$ for the four ensembles (100 states each): (a) EAR packing, (b) hyperuniform fluid ($p_0=4.2, K=1)$, (c) hyperuniform solid ($p_0=3.75, K=0.01$), and (d) non-hyperuniform solid ($p_0=3.75, K=1$). Dashed lines represent the mean. Inset shows the distribution of local $\psi_6$ values in each case.
    }
    \label{fig:orientational_order}
\end{figure}
\clearpage

\begin{figure}[p]
    \centering
    \includegraphics[width=1\columnwidth]{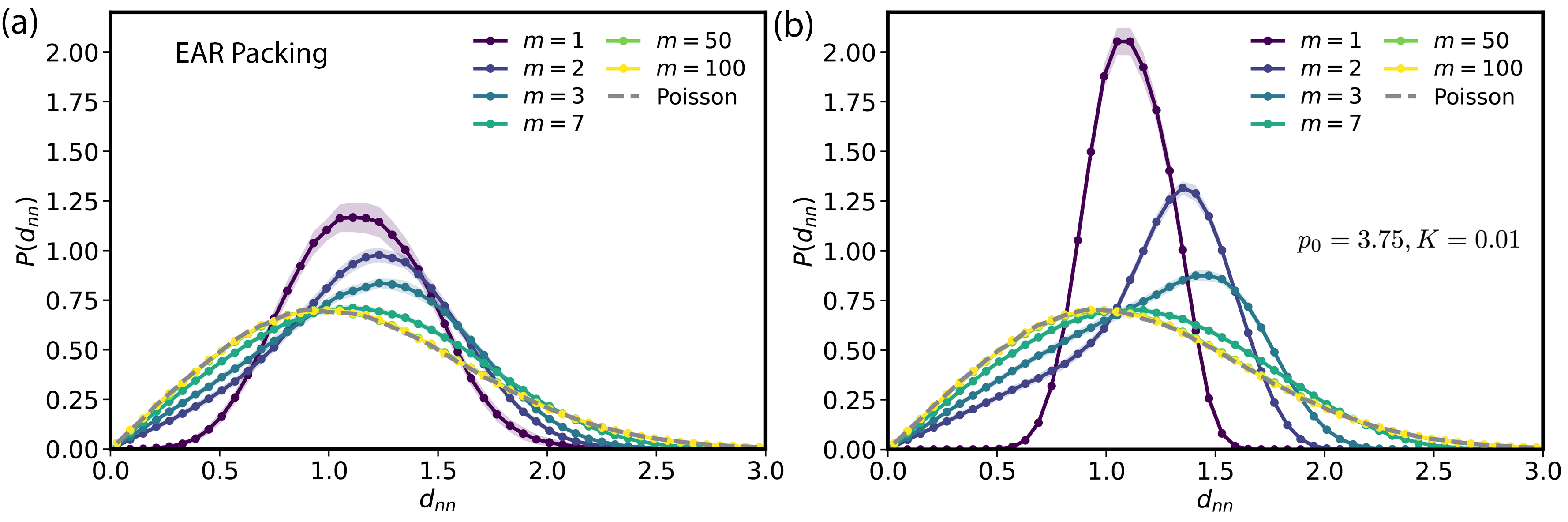}
    \caption{\textbf{Distribution of nearest-neighbor distances $P(d_{\mathrm{nn}})$ for HyPE states.}
    Probability density function of nearest-neighbor distances $d_{\mathrm{nn}}$ between cells for $m$ layers of HyPE states composed of (a) EAR packing layers and (b) hyperuniform solid states with $p_0=3.75$ and $K=0.01$. Shaded bands indicate $\pm 1$ standard deviation over 100 realizations.
    }
    \label{fig:hype_dnn}
\end{figure}

\clearpage

\end{document}